\documentclass[preprintnumbers,twocolumn,aps,nofootinbib]{revtex4}
\pdfoutput=1
\usepackage{amsmath}
\usepackage{amssymb}
\usepackage{bbm}
\usepackage{graphics,graphicx}
\usepackage{dsfont}
\usepackage[raggedright,nooneline]{subfigure}

\newcommand{\mmod}{\text{~mod~}}

\begin{document}
\begin{flushleft}
IFT-UAM/CSIC-17-009
\end{flushleft}

\vspace{1cm}
\title{One-loop Pfaffians and large-field inflation in string theory}

\author{Fabian Ruehle\footnote{fabian.ruehle@physics.ox.ac.uk}}
\affiliation{Rudolf Peierls Centre for Theoretical Physics, Oxford University, 1 Keble Road, Oxford, OX1 3NP, UK}
\author{Clemens Wieck\footnote{clemens.wieck@uam.es}}
\affiliation{Departamento de F\'isica Te\'orica and Instituto de F\'isica Te\'orica UAM/CSIC,\\ Universidad Aut\'onoma de Madrid, Cantoblanco, 28049 Madrid, Spain}

\begin{abstract}

We study the consistency of large-field inflation in low-energy effective field theories of string theory. In particular, we focus on the stability of K\"ahler moduli in the particularly interesting case where the non-perturbative superpotential of the K\"ahler sector explicitly depends on the inflaton field. This situation arises generically due to one-loop corrections to the instanton action. The field dependence of the modulus potential feeds back into the inflationary dynamics, potentially impairing slow roll. We distinguish between world-sheet instantons from Euclidean D-branes, which typically yield polynomial one-loop Pfaffians, and gaugino condensates, which can yield exponential or periodic corrections. In all scenarios successful slow-roll inflation imposes bounds on the magnitude of the one-loop correction, corresponding to constraints on possible compactifications. While we put a certain emphasis on Type IIB constructions with mobile D7-branes, our results seem to apply more generally.

\end{abstract}

\maketitle

\tableofcontents

\section{Introduction}
\label{sec:Introduction}

Cosmic inflation is the leading paradigm to explain the anisotropies of the Cosmic Microwave Background (CMB) radiation. As inflationary theories---especially those involving super-Planckian field excitations---are highly UV sensitive, much work has been devoted to describing those theories within string compactifications. However, a study of the cosmological history in phenomenologically realistic string constructions with time-dependent four-dimensional backgrounds is technically challenging. Therefore, one usually addresses the inflationary dynamics by means of low-energy effective field theories (EFTs). Typically such an EFT has multiple cut-off scales, like the mass scales of string and Kaluza-Klein excitations and the mass scales of geometric moduli. Hence, the high energy density involved in large-field inflation necessitates a careful consistency check of all EFT limits.

Fortunately, many important aspects of this consistency can be studied in low-energy effective supergravities of different string compactifications. We wish to focus on one aspect in particular, the consistency of integrating out heavy K\"ahler moduli during large-field inflation. In Type IIB flux compactifications, the stabilization of the K\"ahler sector is particularly well understood \cite{Giddings:2001yu,Kachru:2003aw}.\footnote{While we focus here on the KKLT stabilization mechanism, we expect our results to be valid in other stabilization schemes as well, such as the ones in \cite{Balasubramanian:2004uy,Kallosh:2004yh,Balasubramanian:2005zx,Conlon:2005ki,Westphal:2006tn,Buchmuller:2016dai}, as shown in \cite{Buchmuller:2015oma}.} The interplay between heavy K\"ahler moduli and large-field inflation in a different sector of the theory has been systematically studied in \cite{Davis:2008fv,Kallosh:2011qk,Buchmuller:2014vda,Buchmuller:2014pla,Buchmuller:2015oma,Dudas:2015lga,Bielleman:2016olv}. Here we wish to extend and generalize those analyses by including one-loop corrections to the non-perturbative superpotential in the K\"ahler sector, which---depending on the details of the microscopic construction---can depend on the inflaton field. This dependence, in turn, feeds back into the inflationary potential once the K\"ahler moduli are stabilized by these non-perturbative terms. Since in a phenomenologically viable model all moduli are stabilized we will focus in this paper on cases where the stabilizing non-perturbative term is non-vanishing at the end of inflation.

We are particularly interested in toy models with a single K\"ahler modulus $T$ and an inflaton multiplet $\Phi$ from a different sector of the theory. Let us consider cases where the low-energy effective superpotential contains at least a constant piece, a quadratic term for $\Phi$, and a non-perturbative piece to stabilize $T$. I.e., we consider\footnote{Note that, with this superpotential, microscopic setups that do not feature a shift symmetry for $\Phi$ do not lead to successful models of inflation. However, our results do not depend on the precise form of $W(\Phi)$, as long as the inflationary EFT features super-Planckian field excitations.}
\begin{subequations}\label{eq:ansatz1}
\begin{align}
K &= K(T+\bar T, \Phi + \bar \Phi)\,,\\
W &= W_0 + \mu \Phi^2 + A (\Phi) e^{-\alpha T}\,.
\end{align}
\end{subequations}
A low-energy effective theory like this can arise as an effective theory of various string compactifications. 

One example are Type IIB flux compactifications with mobile D7-branes \cite{Ibanez:2014swa,Bielleman:2016grv,Bielleman:2016olv}. There $T$ parameterizes the volume of a four-cycle in the compact manifold $Y_6$ and $\Phi$ is the position modulus of a single mobile D7-brane, also wrapping a four-cycle. Moreover, $W_0$ and $\mu$ are sourced by components of $G_3$ flux on $Y_6$ and the non-perturbative term is sourced, for example, by a gaugino condensate on a separate stack of D7-branes. In that case $\alpha$ depends on the rank of the condensing gauge group. Another interesting Type IIB construction with an effective theory like \eqref{eq:ansatz1} is given in \cite{Hebecker:2014eua,Arends:2014qca}. There the $\mu$-term for the inflaton is sourced by couplings to other heavy moduli which are already integrated out.\footnote{See, however, \cite{Hebecker:2014kva} for a discussions of the problems that arise when $\mu$ is tuned to a small value.} 

In heterotic string/M-Theory and its F-Theory lift, couplings like the ones of \eqref{eq:ansatz1} can arise, for example, from world-sheet instantons. In this case $\Phi$ is a bundle or complex structure modulus, $\alpha$ is given in terms of the Gromov-Witten invariants of the corresponding cycle and $\mu$ and $W_0$ depend on the expectation values of other heavy fields.

In \cite{Buchmuller:2015oma,Bielleman:2016olv} a Type IIB toy model with KKLT stabilization like the one in \eqref{eq:ansatz1} has been extensively studied to understand the backreaction of $T$ on the inflationary dynamics. For constant $A$, which is assumed in those references, the result is that, as long as $W_0$ and $\mu$ can be tuned independently, a large mass hierarchy between inflaton and modulus can lead to the possibility of 60 or more $e$-folds of slow roll inflation in agreement with the most recent CMB data. However, this ansatz implies a few non-trivial simplifications. First, the arrangement of the aforementioned four-cycles is drastically simplified, since \eqref{eq:ansatz1} only contains a single four-cycle volume parameter. We comment on more complicated cases later on. Second, all other fields and moduli, especially the complex structure and dilaton, are assumed to be stabilized at a high scale and play no role in the discussion, which seems to be a reasonable assumption~\cite{Bielleman:2016olv}. Third, depending on the details of the microscopic construction, the coefficient $A$ in $W$ may depend on the open-string field $\Phi$ through one-loop corrections to the instanton action. The precise form of $A(\Phi)$ is determined by the details of the geometry and the microscopic setup that leads to an inflationary universe. Technically, $A(\Phi)$ can be found by computing the Pfaffian of the one-loop diagram correcting the instanton action, which is a notoriously difficult task. This computation has only been performed explicitly in a few examples.

The most important distinction is the origin of the instanton in question. We consider the two most common cases, world-sheet instantons and gaugino condensates. For the latter, in Type IIB $A(\Phi)$ arises as open-string one-loop corrections to the gauge kinetic function on the four-cycle parameterized by $T$ \cite{Berg:2004ek,Berg:2004sj},
\begin{align}
 f = \alpha T + \frac{1}{4\pi^2} \log[g(\Phi)] + \dots\,,
\end{align}
where $g$ also depends on expectation values of complex structure moduli. However, none of the examples where $g$ is known explicitly features all ingredients (such as $G_3$ flux) necessary to realize large-field inflation. The one-loop Pfaffian for Euclidean D-brane instantons, as a function of brane position moduli, has been studied in Type II/F-theory in \cite{Blumenhagen:2006xt,Ibanez:2006da,Florea:2006si,Cvetic:2012ts,Corvilain:2016kwe}, cf.~also the earlier discussions in \cite{Ganor:1996pe,Giddings:2005ff}. In heterotic string theory the relevant references include \cite{Dine:1986zy,Dine:1987bq,Buchbinder:2002ic,Curio:2008cm,Curio:2009wn}. Here the Pfaffians have been shown to yield homogeneous polynomials in complex structure and bundle moduli. Furthermore, the authors of \cite{Baumann:2006th,Baumann:2007ah,McAllister:2016vzi} have shown that for both types of instantons the one-loop correction, in terms of brane moduli, can be found using purely closed-string methods. These can be used even in more complicated setups than those considered in \cite{Berg:2004ek,Berg:2004sj}. Moreover, in \cite{Grimm:2007hs,Grimm:2014vva,Higaki:2015kta,Higaki:2016ydn} the authors study setups where one-loop Pfaffians generate the leading-order inflaton potential.  

The rest of this paper is structured as follows. In Section~\ref{sec:Pol} we discuss the possibility that $A(\Phi)$ is a polynomial function. We discuss the string theory backgrounds that produce the corresponding low-energy EFT, and address the consistency of integrating out $T$ in these cases. In the resulting inflationary EFT we derive parameter bounds that constrain possible compactifications and depend on the degree of the polynomial. In Section~\ref{sec:ExpPer} we discuss the possibility that $A(\Phi)$ is exponential or periodic, cases known to arise in Type IIB models with gaugino condensates and mobile D7-branes. We repeat the same analysis as in the polynomial case to study the viability of inflation. Moreover, we hint at interesting CMB signatures induced by modulations of the inflaton potential through the one-loop Pfaffian. Finally, we conclude in Section~\ref{sec:Conclusion}.

\section{Polynomial Pfaffians}
\label{sec:Pol}

\subsection{Motivation}

Let us start by motivating how polynomial Pfaffians arise in string theory, following \cite{Donagi:2010pd,Cvetic:2012ts}. In the case of Type IIB, such instantons arise from Euclidean D3 (ED3) branes that wrap an internal four-cycle. In the M-theory lift, this corresponds to Euclidean M5-branes wrapping vertical divisors that are elliptic fibrations over a divisor in the base of a Calabi-Yau four-fold. In the corresponding F-theory description, the elliptic fiber is shrunk and the Euclidean M5-branes correspond to ED3-branes. Non-perturbative superpotentials arise from the intersection of the four-cycle wrapped by the ED3-branes with the GUT divisor. The moduli $\Phi$ are ED3- and D7-brane moduli in this setup, and here we focus on the D7-brane moduli. If a heterotic dual exists, the Pfaffians arise from world-sheet instantons on a curve whose volume is governed by the K\"ahler parameter $T$. The relevant moduli $\Phi$ are in this case bundle and complex structure moduli, cf.\ \cite{Buchbinder:2002ic,Curio:2008cm,Curio:2009wn,Cvetic:2012ts}. The one-loop Pfaffian is then a homogeneous polynomial (or a product of homogeneous polynomials) to some power. In the examples of \cite{Buchbinder:2002ic,Cvetic:2012ts} there is a single polynomial to the fourth power, and it was speculated that the power to which the polynomial is raised is linked to the number of zero modes of the fermionic differential operator from which the Pfaffian arises. Interestingly, as we shall see later, the results and implications for inflation differ depending on this power. 

A full analysis of the entire moduli space including all dynamics is, so far, impossible due to a lack of explicit expressions for the K\"ahler potentials and computer power. Hence, we work under the assumption that there is a separation of scales such that some moduli have already been stabilized and integrated out at some higher scale. We treat those as effective constants and minimize the potential in terms of the lightest moduli that remain as dynamical parameters. We thus make the following ansatz for the superpotential,
\begin{align}
\label{eq:Superpotential}
W = W_0+A_0\left(\sum_{m=0}^{2n}\delta_m\Phi^m\right)e^{-\alpha T}+\mu\Phi^2\,,
\end{align}
where the constants $W_0$, $\delta_m$, and $\mu$ of the effective theory are given in terms of products of vacuum expectation values of heavy fields. In the following we use the notation
\begin{align}
\Phi|_{\theta=0}=\frac{1}{\sqrt{2}}(\chi+i\varphi)\,,\qquad T|_{\theta=0}=t+i\sigma\,,
\end{align}
for the lowest components of the chiral superfields. Eventually, we are interested in large-field inflation where the imaginary part $\varphi$ of $\Phi$ serves as the inflaton. That is why we have singled out the term $\mu\Phi^2$, which sources the inflaton potential and stabilizes the field in the vacuum. As mentioned in the introduction, a $\mu$-term is only an example and our qualitative results apply more generally. While it is clear that $\mu$, in this case, must depend on expectation values of heavy fields, we are not interested in the details of its origin. Instead, we focus on the impact of the nontrivial one-loop correction in cases where a quadratic term does arise. Moreover, in situations with multiple zero modes where the Pfaffian is given in terms of a polynomial to some power $p$, we have $\delta_m=0$ for $m \mmod p\neq0$, but we keep the discussion general here, since it is easy to specialize later. 

The second ingredient we need in order to study the scalar potential and the inflationary physics derived from it is the K\"ahler potential. The precise form of the K\"ahler potential is not crucial for our discussion as long as it is shift-symmetric in the imaginary part of $\Phi$. We use the following ansatz for the K\"ahler potential,
\begin{align}
\label{eq:KahlerPotential}
K = -3 \log{(T + \bar T)} + \frac12 (\Phi + \bar \Phi)^2\,.
\end{align}
In most string compactifications, like in Type IIB where $\Phi$ is a D7-brane modulus or in heterotic string theories where $\Phi$ is a complex structure modulus\footnote{If $\Phi$ is a bundle modulus the precise form of the K\"ahler potential is unknown.}, one would generally expect a dependence of the form ${K \supset - \log(c+\Phi+\bar\Phi)}$. Since $c$ is a constant in the EFT, determined by expectation values of heavy fields, we can expand this expression around small values of $\text{Re}(\Phi)$ to obtain \eqref{eq:KahlerPotential}.

With this we can calculate the scalar potential in terms of $T$ and $\Phi$. Assuming $\delta_0 \neq 0$, there is a stable AdS vacuum at $\Phi = 0$ and some non-vanishing value of $T$, as in the original setup of KKLT. Thus, we add an uplifting term to the potential of the form $V_\text{up}=e^K\Delta$, which is then tuned to obtain a Minkowski vacuum.\footnote{A term like this can be sourced, for example, by heavy matter fields which break supersymmetry. Assuming the dynamics of the uplifting sector decouple beyond inflation, the precise field-dependence of the uplift does not affect our results.}

\subsection{Backreaction and large-field inflation}

\subsubsection{Generalities}

Generically there is more than one zero mode for a Pfaffian and the homogeneous polynomial is raised to some power $p$. This, in turn, means that all powers of $\Phi$ that appear are divisible by $p$. For this reason we discuss cases where $\delta_m=0$ unless $p$ divides $m$. Moreover, since the $\delta_m$ are, in this case, given by products of expectation values of heavy fields, one expects that $\delta_m\gg\delta_{m+1}$ in a controlled regime of the compactification like the large complex structure regime. This is because the overall Pfaffian polynomial is homogeneous, so that the higher the power of $\Phi$ the lower the number of fields that enter a given superpotential coupling. 

With this in mind we consider only the lowest-order term in the  polynomial \eqref{eq:Superpotential}. We thus work with the K\"ahler potential \eqref{eq:KahlerPotential} and the superpotential
\begin{align}
\label{eq:SuperpotentialOneTerm}
W=W_0+A_0(1+\delta \Phi^m)e^{-\alpha T}+\mu\Phi^2\,.
\end{align}
This allows for a much clearer illustration of our results. The generalization to a sum of different contributions is then straight-forward. Note that, compared to \eqref{eq:Superpotential}, we have absorbed $\delta_0$ in $A_0$ and renamed $\delta_m/\delta_0=:\delta<1$. The scalar potential of the EFT thus depends on the fields $\Phi$ and $T$, and the parameters $W_0$, $A_0$, $\delta$, $\alpha$, $\Delta$, and $\mu$. 

Using the scalar potential $V(T,\Phi)$ we can now study whether large-field inflation driven by $\varphi$ is still possible and the modulus $T$ remains under control. For the sake of concreteness we choose the following parameters in all examples,
\begin{align}
\label{eq:ParameterExamplesPoly}
\alpha=\frac{2\pi}{5}\,,\quad W_0=4\times10^{-3}\,,\quad \mu=10^{-5}\,.
\end{align}
Furthermore, by demanding the existence of a Minkowski vacuum at $\varphi = 0$ we can eliminate the two parameters $A_0$ and $\Delta$ in terms of the ones in \eqref{eq:ParameterExamplesPoly} and $t_0$, the position of the true vacuum. With $t_0 = 10$ one finds $A_0=-\mathcal{O}(100)$ and $\Delta=\mathcal{O}(W_0)$.

Notice that setting $\delta = 0$ reproduces the results of \cite{Buchmuller:2015oma}. There it was shown that, since $W_0$ determines the mass scale of $T$ and $\mu$ determines the mass scale of $\Phi$, if $W_0 \gg \mu$ the modulus can be integrated out to yield an effective inflaton potential. In \cite{Buchmuller:2015oma} it sufficed to consider the potential to quadratic order in $t$. Expanding around the post-inflationary vacuum, $t = t_0 + \Delta t(\varphi)$ and minimizing the result yielded the correct leading-order effective potential for the inflaton. This, unfortunately, is no longer true for our more complicated scalar potential. First, terms with odd $m$ lead to a displacement of $\sigma$ during inflation, in addition to that of $t$. Thus, $\sigma$ must be integrated out as well to obtain the correct effective potential. Second, as we will see below, the potential of $t$ is too complicated to be approximated by a second-order expansion around the minimum $t_0$. Since solving higher-order equations of motion is technically challenging, we resort to numerical methods to determine the inflationary potential with sufficient accuracy. 

\subsubsection{Parameter examples}

\begin{figure*}[t]
 \subfigure[][~Scalar potential for $m=1$ in the $t$-$\varphi$-plane.]{
    \includegraphics[width=.88\columnwidth]{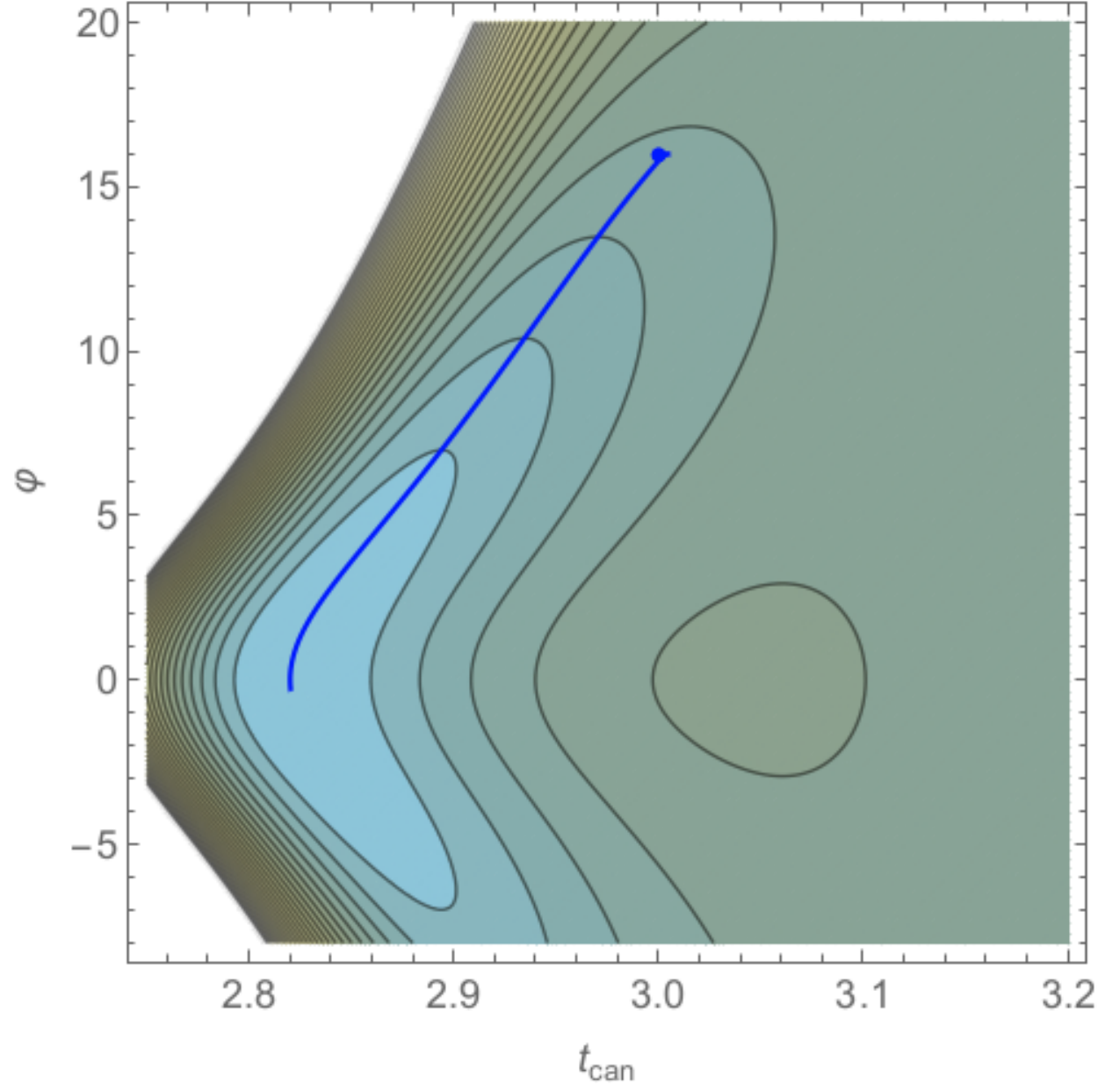}
    \label{fig:cont1}
 }\qquad\quad
 \subfigure[][~Effective scalar potential for $m=1$ on the blue trajectory.]{
     \raisebox{9mm}{\includegraphics[width=.91\columnwidth]{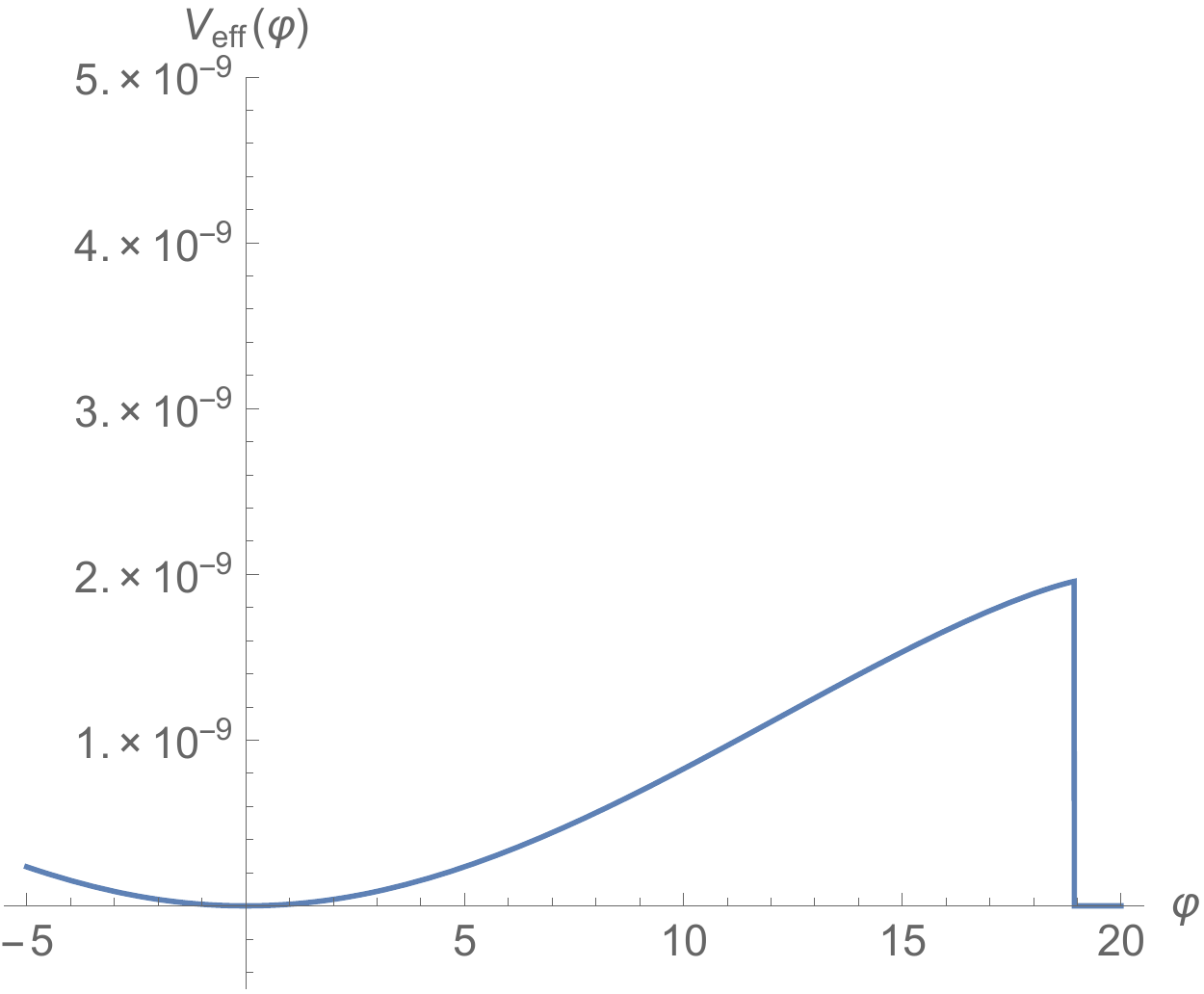}}
     \label{fig:Veff1}
 } 
 \caption{\subref{fig:cont1} Full scalar potential $V(t,\varphi)$ for $m = 1$ and $\delta = 1/3$ after numerically integrating out $\sigma(\varphi,t)$. The blue line illustrates a possible inflationary trajectory, ending in the metastable Minkowski vacuum. As $\delta t $ is positive in this case, the inflationary valley turns to the right until it disappears at $\varphi \approx 19$. \subref{fig:Veff1} Effective scalar potential on the inflationary trajectory. $t (\varphi)$ and $\sigma (\varphi)$ are minimized numerically. The point where the curve drops is the point where there is no longer a local minimum in the $t$ direction. A universe with initial conditions beyond this point would evolve towards decompactification at $t \to \infty$, a most undesirable situation.}
 \label{fig:Case1}
\end{figure*}

\begin{figure*}[t]
 \centering
 \subfigure[][~Scalar potential for $m=2$ in the $t$-$\varphi$-plane.]{
    \includegraphics[width=.88\columnwidth]{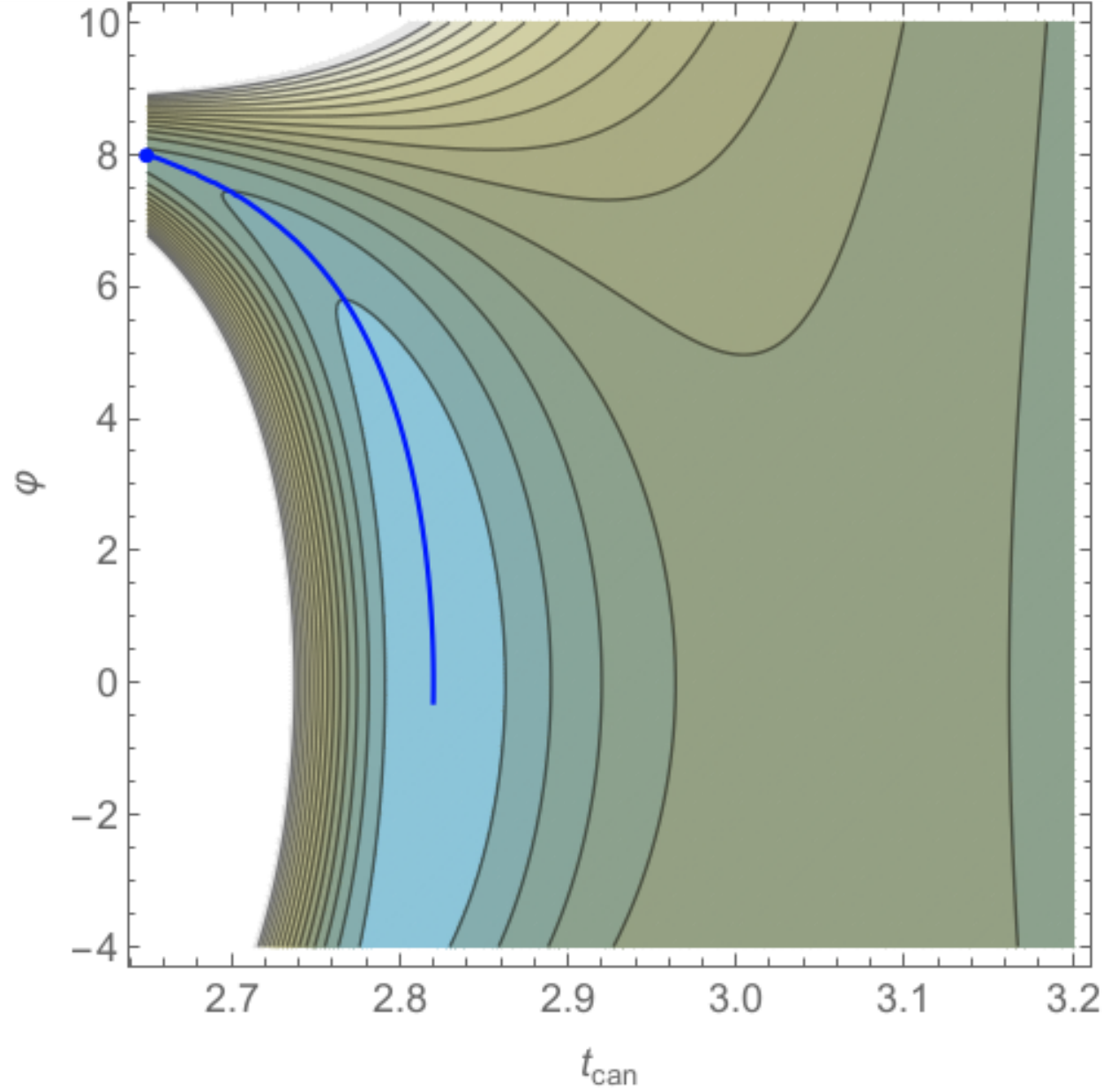}
    \label{fig:cont2}
 }\qquad\quad
 \subfigure[][~Effective scalar potential for $m=2$ on the blue trajectory.]{
      \raisebox{9mm}{\includegraphics[width=.91\columnwidth]{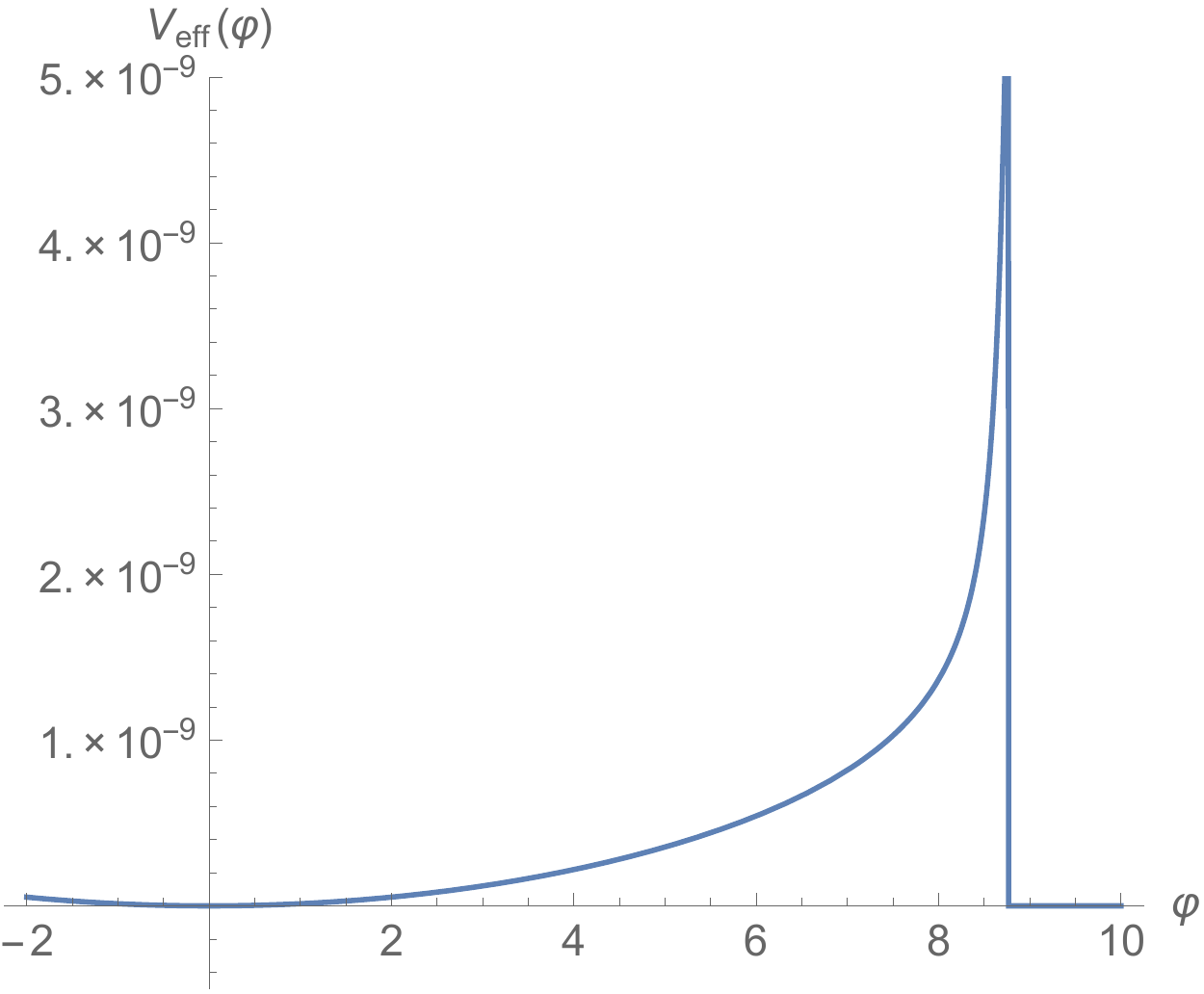}}
     \label{fig:Veff2}
 } 
 \caption{\subref{fig:cont2} Full scalar potential $V(t,\varphi)$ for $m = 2$ and $\delta = 1/4 \pi^2$ after numerically integrating out $\sigma(\varphi,t)$. The blue line illustrates a would-be inflationary trajectory, ending in the metastable Minkowski vacuum. As $\Delta t(\varphi)$ is negative in this case, the inflationary valley turns to the left towards the exponentially steep flank. This happens at $\varphi \sim 1/\sqrt \delta$. \subref{fig:Veff2} Effective scalar potential on the blue trajectory. $t (\varphi)$ and $\sigma (\varphi)$ are minimized numerically.}
 \label{fig:Case2}
\end{figure*}

There is, however, one observation which can already be made at the level of the quadratic approximation: the different terms in the sum in \eqref{eq:Superpotential} have different effects depending on whether $m = 2n + 1$, $m = 4n + 2$, or $m = 4n$ with $n\in\mathbbm{Z}^+$. The structure of the inflationary trajectory is greatly affected by the sign of the first-order term in the expansion around $t_0$. If it is negative, so that $\Delta t(\varphi) < 0$ to leading order, the inflationary trajectory becomes exponentially steep for large inflaton field values. If it is positive, so that $\Delta t(\varphi) > 0$ to leading order, the trajectory is flat but the modulus may be destabilized for too large field values. To see this explicitly, consider the expressions in the relevant expansion,
\begin{align}
\label{eq:ScalarPotentialApprox}
V(t, \varphi) = V_0 + (t - t_0) V_1 + (t - t_0)^2 V_2 + \dots \,,
\end{align}
where
\begin{align}
V_0 	&= \frac{1}{32t_0^3}\Big\{16 \mu^2 \varphi^2  + 6  \delta^2 W_0^2 \Big(\frac{\varphi^{2}}{2}\Big)^m \nonumber\\
        	&~+\, 6 \mu W_0 \varphi^2 \left[1 + 2^{-\frac{m+2}{2}} \left((-i \varphi)^m + (i \varphi)^m \right) \delta\right]\Big\}\,,\label{eq:v0}\\
V_1 	&= -\frac{6\alpha W_0}{32 t_0^3}\Big\{ \mu \varphi^2 +2\delta^2W_0 \Big(\frac{\varphi^{2}}{2}\Big)^m\nonumber\\
        	& \quad+ 2^{-\frac{m+2}{2}} \left((-i \varphi)^m + (i \varphi)^m \right)\left[\mu\varphi^2+2W_0\right]\delta \Big\}\,, \label{eq:v1}\\
V_2 	&= \frac{3\alpha^2 W_0}{32t_0^3}\Big\{ \mu \varphi^2 + 2W_0+4\delta^2W_0 \Big(\frac{\varphi^{2}}{2}\Big)^m\nonumber\\
       	& \quad +2^{-\frac{m+2}{2}} \left((-i \varphi)^m + (i \varphi)^m \right)\left[\mu\varphi^2 + 6W_0\right] \delta\Big\}\,,\label{eq:v2}
\end{align}
to lowest order in inverse powers of $t_0 \gg 1$. Moreover, for the sake of simplicity we have set $\sigma = 0$, which does not affect the following arguments, as we have checked a posteriori. 
 
\begin{center}
$\boldsymbol{m=1 \textbf{\mmod} 2}$
\end{center}

Clearly, if $m$ is odd, $V_1$ in \eqref{eq:v1} is negative for any value of $\delta$ and $\varphi$. Furthermore, $V_2$ in \eqref{eq:v2} is positive for any reasonable choice of parameters since it is proportional to the squared mass of the modulus. Hence $\Delta t(\varphi) = -V_1/2 V_2 + \dots$ is positive. That this is true to all orders in $t$ and $\sigma$ is displayed in Figure~\ref{fig:cont1}. It features the full scalar potential in the $t$-$\varphi$-plane for $m=1$ and $\delta = 1/3$, in terms of the canonically normalized modulus denoted by $t_\text{can}$. No approximations have been made and $\sigma$ is integrated out dynamically at its field-dependent minimum. There is a stable valley flat enough to support slow-roll inflation ending at $\varphi = 0$, indicated by the blue trajectory which gives the numerical solution to the full equations of motion of the system. Although the inflationary evolution proceeds along a curved line in the $t$-$\varphi$-plane, this is still single-field inflation: the orthogonal direction to the blue line is very steep, due to the hierarchy $W_0 \gg \mu$.

The modulus is destabilized once the inflaton field excursion becomes too large, in this case at $\varphi \approx 19$. To further illustrate this point, Figure~\ref{fig:Veff1} contains the effective potential in the flat valley, obtained after numerically integrating out the heavier direction. As compared to the naive case with $\delta = 0$, the inflationary valley is tilted to the right, towards the KKLT barrier and run-away. However, the point of destabilization of the modulus is, as in \cite{Buchmuller:2015oma}, still only affected by the ratio $W_0/\mu$. In particular, it does not depend on $\delta$. Moreover, the potential is always flat as long as the valley exists. This means that the one-loop dependence of $A(\Phi)$ can, for odd $m$, be quite significant without endangering 60 $e$-folds of slow-roll inflation. Only when $\delta$ is large, successful inflation is impossible due to the appearance of a false minimum on the flat trajectory. This happens when $\delta \gtrsim (\varphi_\star/10)^{-m}$, where $\varphi_\star \sim 15$ is the starting point of the last 60 $e$-folds of slow-roll inflation.

In the example chosen above, 60 $e$-folds of slow-roll inflation may take place on the blue trajectory. We find the following predictions for the scalar spectral index and the tensor-to-scalar ratio,
\begin{align}
n_\text s \approx 0.964 \,, \quad r \approx 0.078\,,
\end{align}
which agree well with the most recent CMB data.

\begin{center}
$\boldsymbol{m=2 \textbf{\mmod} 4}$
\end{center}

If $m = 4n + 2$, $V_1$ is positive for most values of $\varphi$ and $\delta$. This changes the picture obtained in the first case dramatically. $\Delta t(\varphi)$ is negative and the inflationary trajectory descends from the exponentially steep flank towards $t\to 0$, instead of the flat part of the potential towards $t \to \infty$. Once more, we have displayed the full scalar potential in the $t$-$\varphi$-plane for $m=2$ and $\delta = 1/4\pi^2$ in Figure~\ref{fig:cont2}. There is again a stable valley towards the metastable Minkowski minimum, but this time it is exponentially steep. If the one-loop Pfaffian is quadratic, $\delta$ must be much smaller for inflation to occur outside of the exponentially steep regime.

\begin{figure*}[t]
 \centering
\subfigure[][~Scalar potential for $m=4$ in the $t$-$\varphi$-plane.]{
    \includegraphics[width=.88\columnwidth]{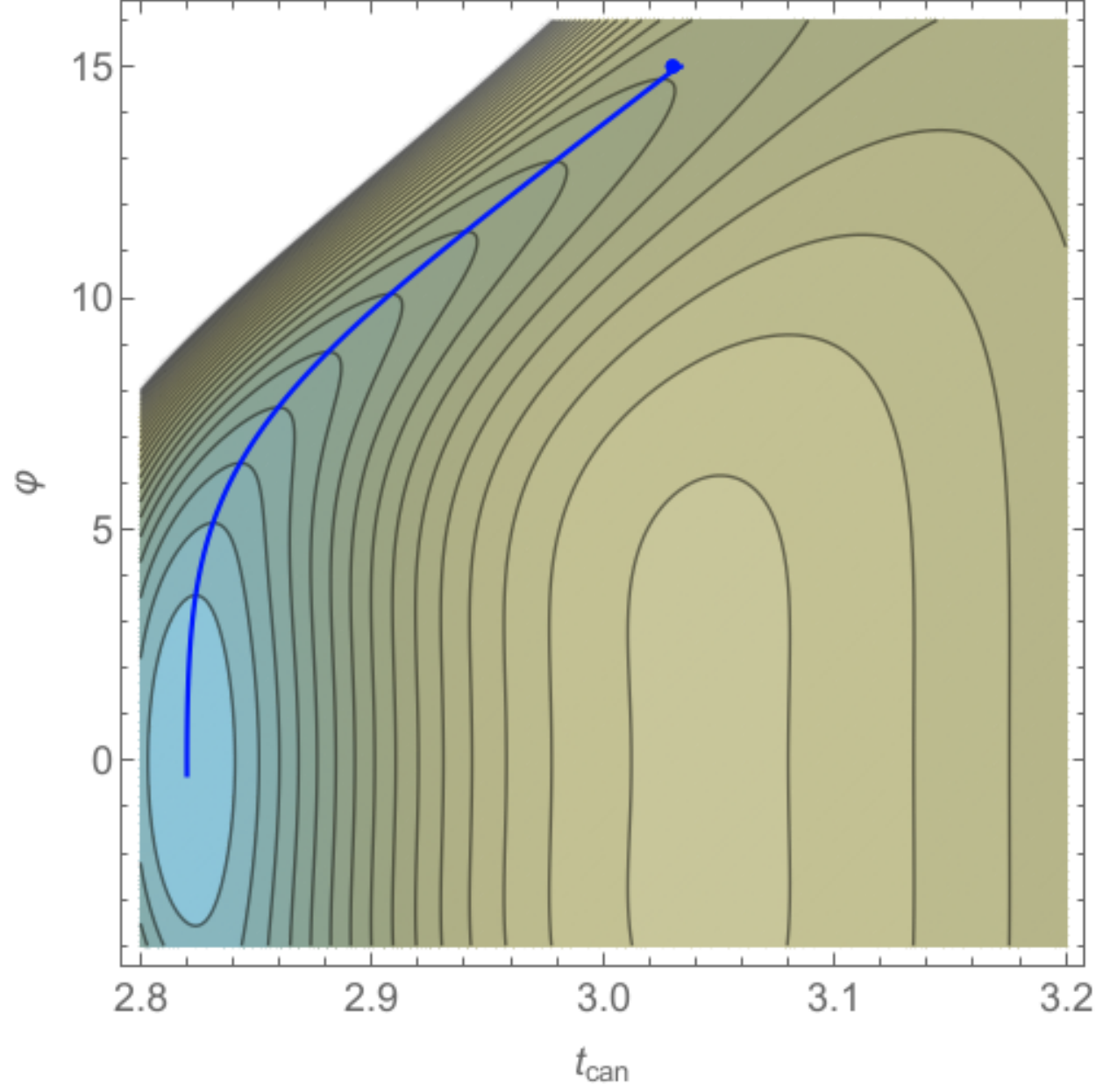}
    \label{fig:cont3}
 }\qquad\quad
 \subfigure[][~Effective scalar potential for $m=4$ on the blue trajectory.]{
      \raisebox{9mm}{\includegraphics[width=.91\columnwidth]{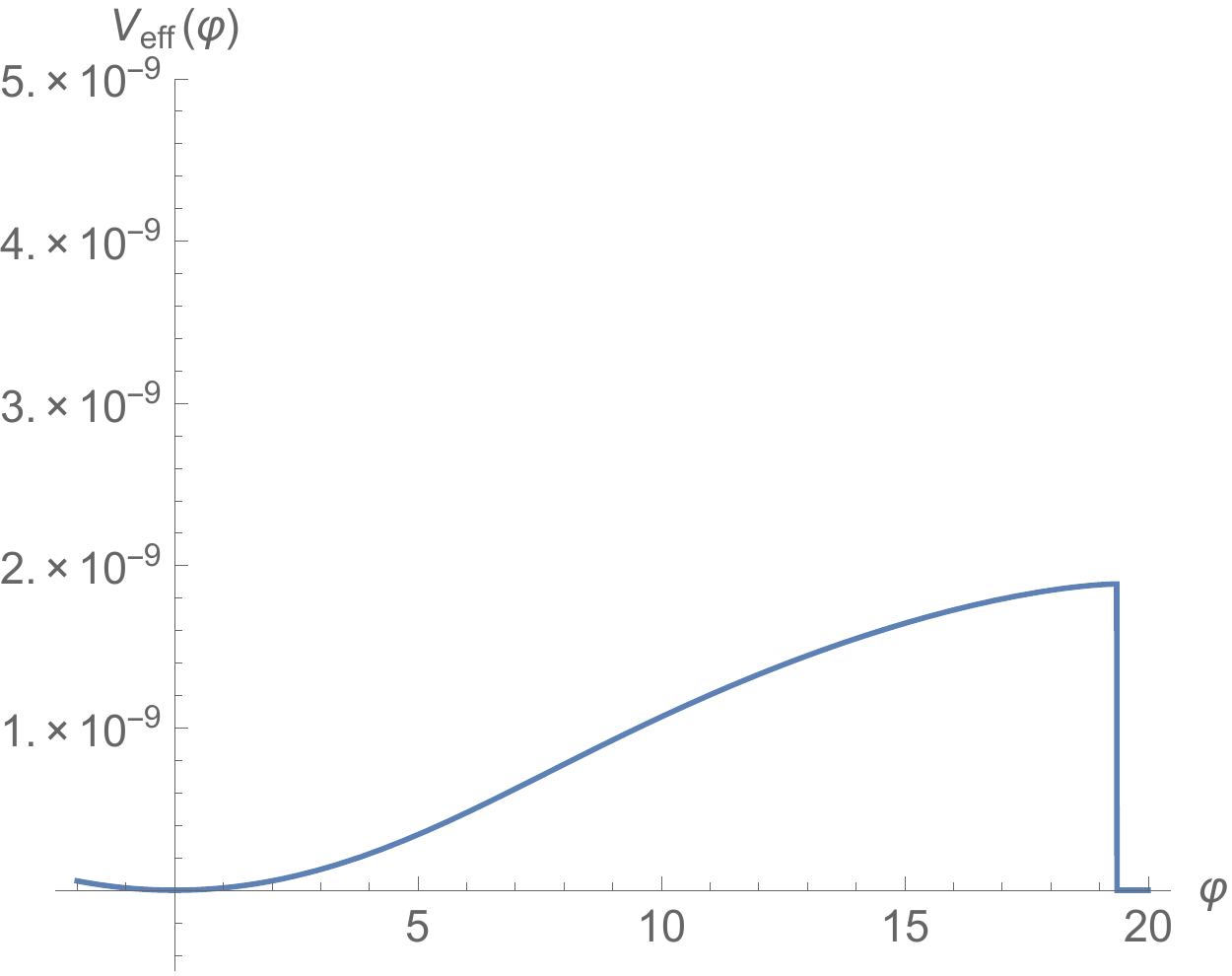}}
     \label{fig:Veff3}
 } 
 \caption{\subref{fig:cont3} Full scalar potential $V(t,\varphi)$ for $m = 4$ and $\delta = 1/256 \pi^2$ after numerically integrating out $\sigma(\varphi,t)$. The blue line illustrates a would-be inflationary trajectory, ending in the metastable Minkowski vacuum. As $\Delta t(\varphi)$ is again positive in this case, the inflationary valley turns towards the flat region of the potential. \subref{fig:Veff3} Effective scalar potential on the blue trajectory. $t (\varphi)$ and $\sigma (\varphi)$ are minimized numerically.}
 \label{fig:Case3}
\end{figure*}

While slow-roll inflation may still be possible, the tensor-to-scalar ratio becomes too large for $\delta \gtrsim \varphi_\star^{-m}$. In other words, if $\delta \varphi^m \sim \mathcal O(1)$ the one-loop correction makes the potential too steep. This is illustrated in the plot in Figures \ref{fig:cont2} and \ref{fig:Veff2}. 

\begin{center}
$\boldsymbol{m=0 \textbf{\mmod} 4}$
\end{center}

The final case, $m = 4n$, is again similar to the first one. Due to an alignment of terms in \eqref{eq:v1}, $V_1$ is again negative for most values of $\varphi$ and $\delta$. The corresponding plots for the example $m = 4$ can be found in Figures~\ref{fig:cont3} and \ref{fig:Veff3}. Again, inflation seems possible for rather large values of $\delta$. The inflationary trajectory sharply turns to the right towards the flat region of the potential. Although the turn happens at $\varphi \approx \delta^{-4}$, the strength of the one-loop correction can be quite large without impairing inflation. On the contrary, the contribution of the Pfaffian flattens the potential further, just like in the first case. For our parameters we find the following predictions,
\begin{align}
n_\text s \approx 0.967 \,, \quad r \approx 0.048\,,
\end{align}
again in good agreement with CMB observations. An upper bound on $\delta$ arises again through the appearance of a false minimum on the inflationary trajectory for ${\delta \gtrsim (\varphi_\star/5)^{-m}}$.

\section{Exponential and periodic Pfaffians}
\label{sec:ExpPer}

In Type II theories where the non-perturbative superpotential is sourced by gaugino condensation on stacks of D-branes, the form of one-loop Pfaffians can be different from the cases discussed in the previous section. For the sake of concreteness, let us focus on type IIB flux compactifications with space-time filling D7-branes. Several constructions exist in the literature which discuss the possibility of axion monodromy inflation driven by the vacuum energy of mobile D7-branes wrapping four-cycles in the compact manifold \cite{Hebecker:2014eua,Arends:2014qca,Ibanez:2014kia,Ibanez:2014swa}. As mentioned above, and discussed in more detail in \cite{Ibanez:2014swa,Bielleman:2016grv,Bielleman:2016olv}, a typical low-energy version of such setups can be formulated as follows,
\begin{subequations}
\begin{align}\label{eq:OpenK1}
K &= -3 \log{(T + \bar T)} -\log\left[c - (\Phi + \bar \Phi)^2\right]\,,\\
W &= W_0 + \mu \Phi^2 + A e^{-\alpha T}\,, \label{eq:OpenW1}
\end{align}
\end{subequations}
where $\Phi$ is the position modulus of a single mobile D7-brane wrapping some four-cycle in the compact manifold $Y_6$. $W_0$ and $\mu$ are sourced by components of $G_3$ flux on $Y_6$ and the non-perturbative term is sourced, in this case, by a gaugino condensate on a separate stack of D7-branes. 

As before, we now consider $A$ to be a function of $\Phi$ whose shape depends on the details of the microscopic setup. Following \cite{Berg:2004ek,Berg:2004sj,Baumann:2006th,Baumann:2007ah,McAllister:2016vzi}, in the case of F-term axion monodromy inflation we can distinguish two interesting limiting cases, which we discuss in the following.

\subsection{Exponential Pfaffians}
\label{sec:Exp}

\begin{figure*}[t]
 \centering
\subfigure[][~Scalar potential in the $t$-$\varphi$-plane.]{
    \includegraphics[width=.88\columnwidth]{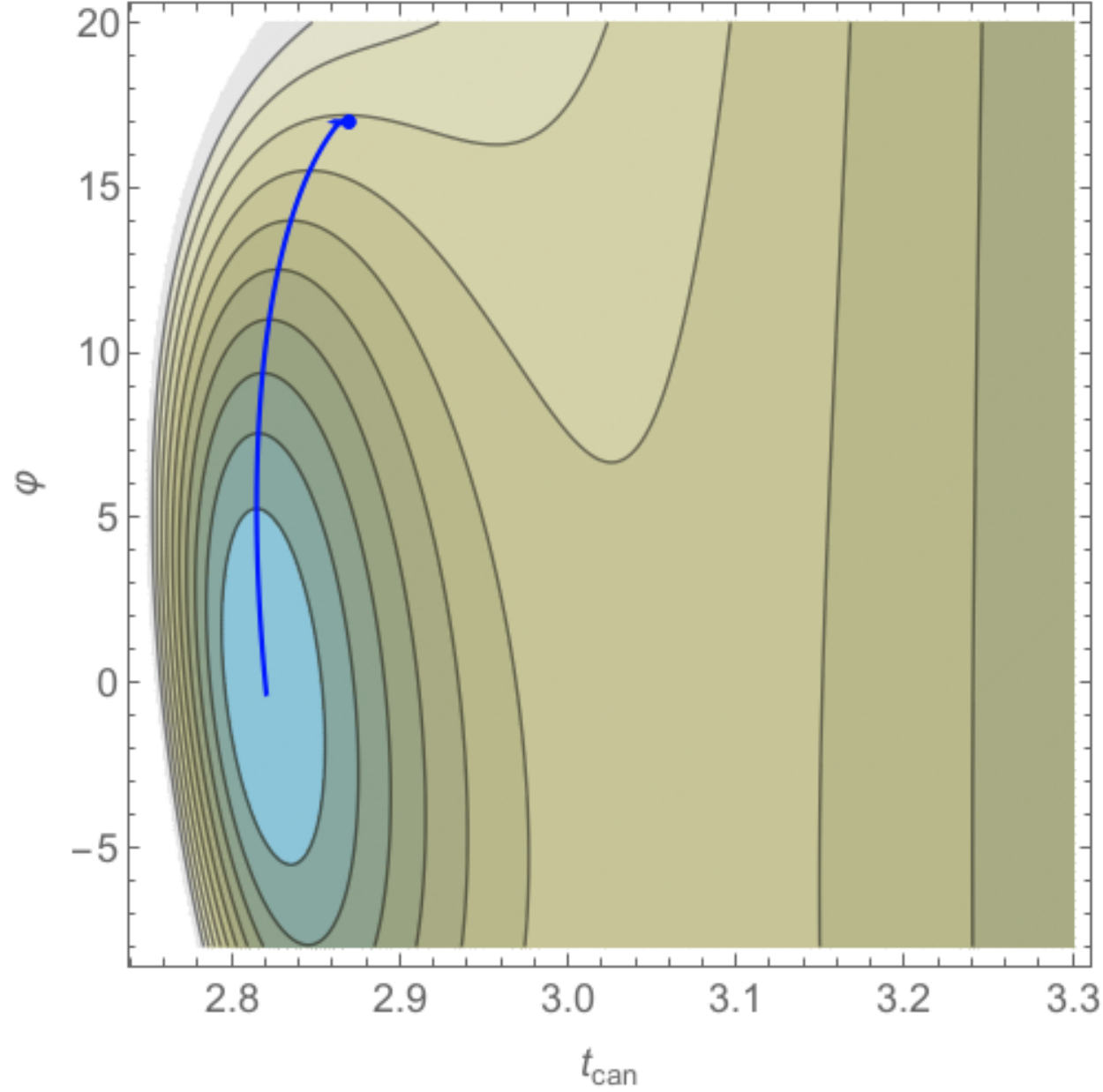}
    \label{fig:Exp1cont}
 }\qquad\quad
 \subfigure[][~Effective scalar potential on the blue trajectory.]{
     \raisebox{7mm}{\includegraphics[width=.91\columnwidth]{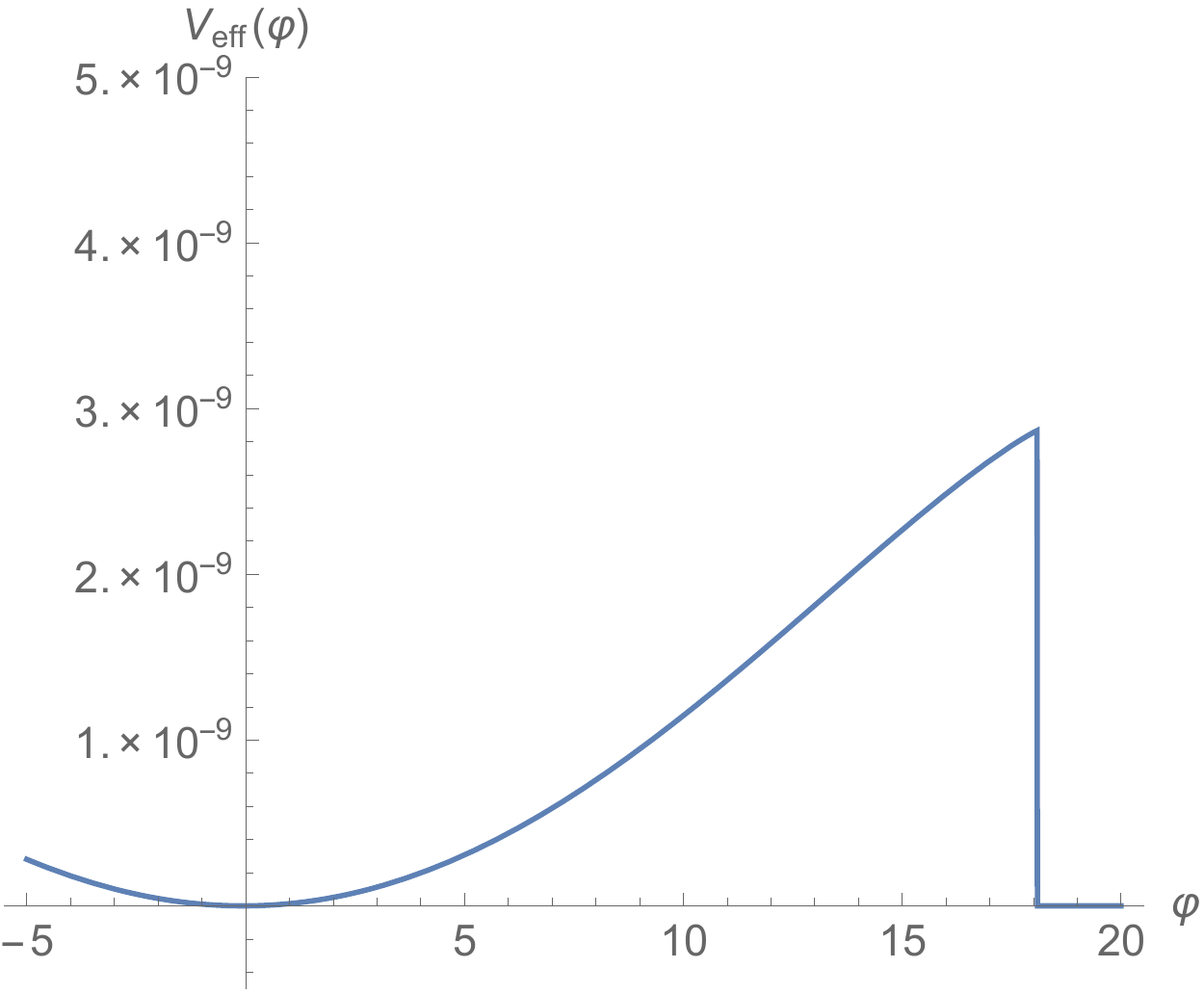}}
     \label{fig:Exp1}
 } 
 \caption{\subref{fig:Exp1cont} Full scalar potential for $\delta = 1/4 \pi^2$ and an exponential Pfaffian. The blue line illustrates a possible inflationary trajectory, ending in the metastable Minkowski vacuum. \subref{fig:Exp1} Effective scalar potential on the blue trajectory. $t (\varphi)$ is minimized numerically. Once more, the point where the curve drops is where $t$ decompactifies. It can be moved to larger values of $\varphi$ by increasing the hierarchy between the parameters $W_0$ and $\mu$.}
 \label{fig:ox1}
\end{figure*}

\subsubsection{Motivation}

The construction of \cite{McAllister:2016vzi} features an axion monodromy potential sourced by an NS5-branes wrapping a two-cycle $\Sigma_2$ in the compact manifold. The induced D3-brane charge on these NS5 branes leads to open strings stretching between the wrapped two-cycle and a four-cycle $\Sigma_4$ supporting a non-perturbative superpotential, just like in our example in \eqref{eq:OpenW1}. Moreover, in that reference the two relevant cycles intersect and are located in a strongly warped region of the internal geometry. In order to realize a large field excursion of the axion $b \sim \int_{\Sigma_2} B_2$ one introduces $N \gg 1$ units of NS5-brane flux, which induces $N\gg1$ units of D3-brane charge on the brane. It was then shown that the large amount of D3-brane charge leads to a strong backreaction on the warped volume of $\Sigma_4$, yielding a dependence of the non-perturbative superpotential on $\Sigma_4$ of the form $A \propto e^{-N}$. Since the axion field value is proportional to the value of $N$, the one-loop Pfaffian depends exponentially on the axion field value.

A similar situation can arise in the D7-brane inflation models discussed above. For instance, in \cite{Ibanez:2014swa} the axion of the D7-brane position modulus has a similar monodromy potential. The D7-brane periodically winds around the two transverse dimensions, and---due to the presence of $G_3$ flux---accumulates with each winding one unit of induced D3-brane charge and an increase in vacuum energy. Therefore, depending on the details of the microscopic setup, the large amount of D3-brane charge can strongly backreact on the four-cycle supporting the gaugino condensate. In particular, this is bound to happen when the four-cycle wrapped by the mobile D7-brane (the one that carries the piled-up D3-brane charge) and the four-cycle with the gaugino condensate (the one parameterized by $T$) intersect or are identified. In this case, we expect the low-energy effective theory to be described by\footnote{Notice that we have again expanded the K\"ahler potential \eqref{eq:OpenK1} around small values of $\text{Re}(\Phi)$, and have absorbed the constant $c$ in the definition of the field.}
\begin{subequations}\label{eq:Open2}
\begin{align}
K &= -3 \log{(T + \bar T)} + \frac12 (\Phi + \bar \Phi)^2\,,\\
W &= W_0 + \mu \Phi^2 + A_0 e^{i \delta \Phi} e^{-\alpha T} \,,
\end{align}
\end{subequations}
where $\delta$ is the one-loop coefficient proportional to $1/4 \pi^2$, which depends on the details of the microscopic setup. In \cite{McAllister:2016vzi} it depends on the local warping and the distance between the different cycles. Here, as in Section \ref{sec:Pol}, we treat it as a constant parameter of the effective theory which can be used to constrain possible geometries and flux choices. In principle, in this kind of setup the one-loop Pfaffian can also have a periodic component, cf.~Section \ref{sec:Per}. However, here the effect of the exponential is much stronger.

\begin{figure*}[t]
 \centering
\subfigure[][~Scalar potential in the $t$-$\varphi$-plane.]{
    \includegraphics[width=.88\columnwidth]{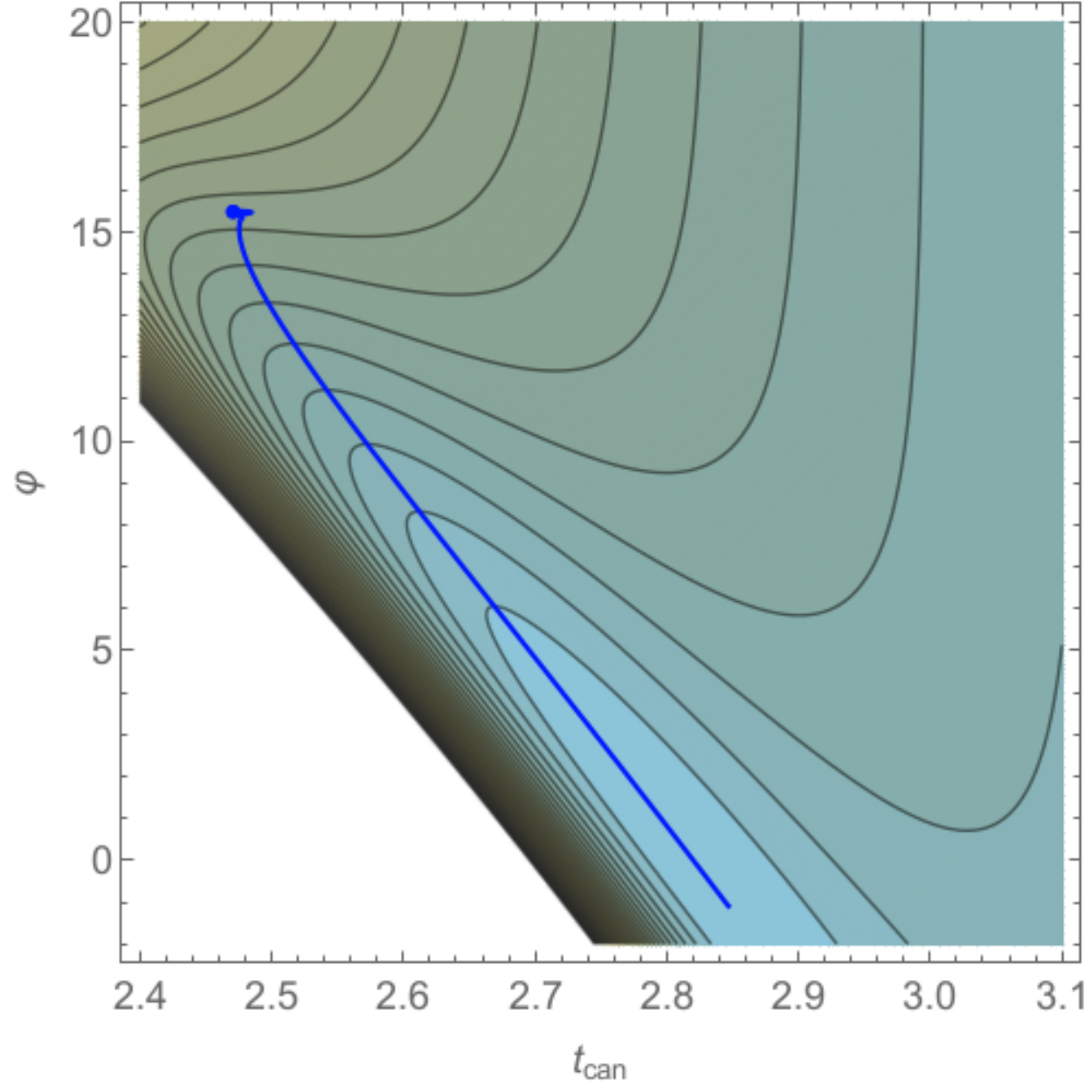}
    \label{fig:Exp2cont}
 }\qquad\quad
 \subfigure[][~Effective scalar potential on the blue trajectory.]{
     \raisebox{7mm}{ \includegraphics[width=.91\columnwidth]{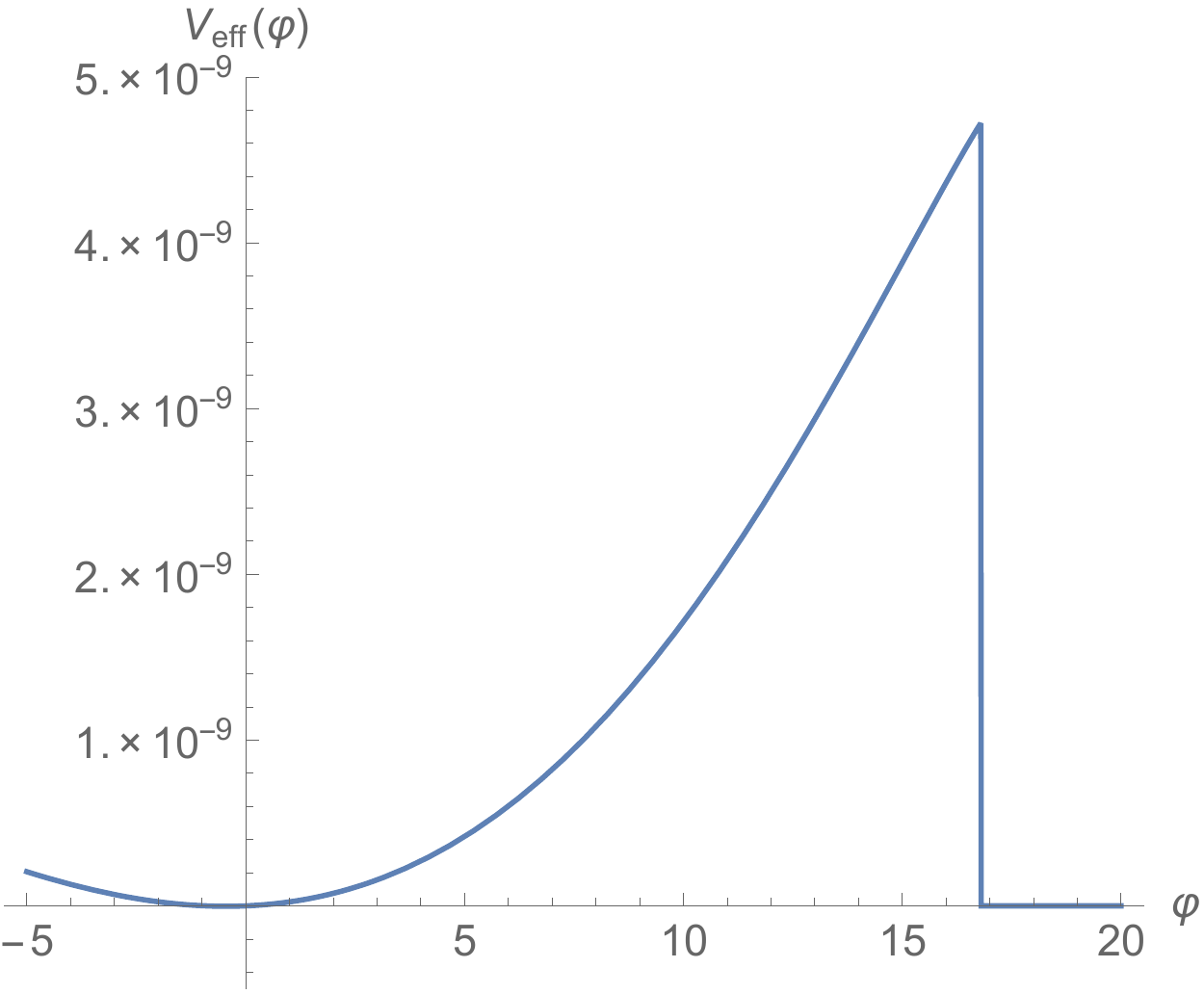}}
     \label{fig:Exp2}
 } 
 \caption{\subref{fig:Exp2cont} Full scalar potential for $\delta = 1/3$ and an exponential Pfaffian. The blue line illustrates a would-be inflationary trajectory, ending in the metastable Minkowski vacuum. \subref{fig:Exp2} Effective scalar potential on the blue trajectory. $t (\varphi)$ is minimized numerically. The potential is too steep to support slow-roll inflation in agreement with observation.}
 \label{fig:ox2}
\end{figure*}

\subsubsection{Large-field inflation and parameter constraints}

We can now repeat the analysis of Section \ref{sec:Pol} for scalar potential defined by \eqref{eq:Open2} and study if large-field inflation is possible in this case. At first sight, the situation seems desperate: as in \cite{McAllister:2016vzi} we need large axion field values $\varphi \gg 1$ which implies a strong suppression of the non-perturbative term, which is responsible for the stability of the K\"ahler modulus. Due to the exponential dependence the effect seems much stronger than in the polynomial case discussed before. However, the field values necessary for chaotic inflation are not nearly as large as those in the relaxation mechanism of \cite{McAllister:2016vzi}, even a moderately small value of the constant $\delta$ may allow for inflation in certain regimes.

Thus, again we have to study the scalar potential in detail. After observing that both the axion $\sigma$ of $T$ and the saxion $\chi$ of $\Phi$ are stabilized at the origin with a large mass and decouple from the inflationary trajectory, we may focus on the potential in the $t$-$\varphi$-plane. As in \cite{Buchmuller:2015oma,Bielleman:2016olv}, a crucial ingredient is still a substantial hierarchy $W_0 \gg \mu$ to ensure stability of $T$ throughout the inflationary epoch.  As a first parameter example, we may choose $t_0 = 10$, $\delta = 1/4 \pi^2$, as well as the parameters in \eqref{eq:ParameterExamplesPoly}. Figure~\ref{fig:Exp1cont} features a contour plot of the scalar potential as a function of $\varphi$ and the canonically normalized modulus field. The metastable Minkowski minimum is clearly visible, as well as a possible inflationary trajectory. The effective scalar potential on this trajectory is displayed in Figure~\ref{fig:Exp1}. It is again obtained after minimizing $t$ as a function of $\varphi$. The CMB observables for 60 $e$-folds of slow roll on this trajectory read
\begin{align}
n_\text s = 0.96\,, \quad r = 0.09\,.
\end{align}

As a second example, let us consider a scenario where $\delta$ is larger and the backreaction is stronger. Choosing the same parameters as above, except $\delta = 1/3$, leads to the scalar potential displayed in Figure~\ref{fig:Exp2cont}. Apparently, increasing $\delta$ leads to a tilt of the flattest direction in the potential. Since the tilt is towards the exponentially steep flank at $t \to 0$, the would-be inflationary trajectory becomes steeper and steeper. We have displayed the corresponding effective potential on the blue trajectory in Figure~\ref{fig:Exp2}. In this particular case the backreaction from the D3-brane charge is too strong: at best, with somewhat fine-tuned initial conditions, one can obtain 50 $e$-folds of slow roll with the following predictions,
\begin{align}
n_\text s = 0.94\,, \quad r = 0.18\,,
\end{align}
a result which is ruled out by observations. 

\subsection{Periodic Pfaffians}
\label{sec:Per}

\begin{figure*}[t]
 \centering
\subfigure[][~Scalar potential in the $t$-$\varphi$-plane.]{
    \includegraphics[width=.88\columnwidth]{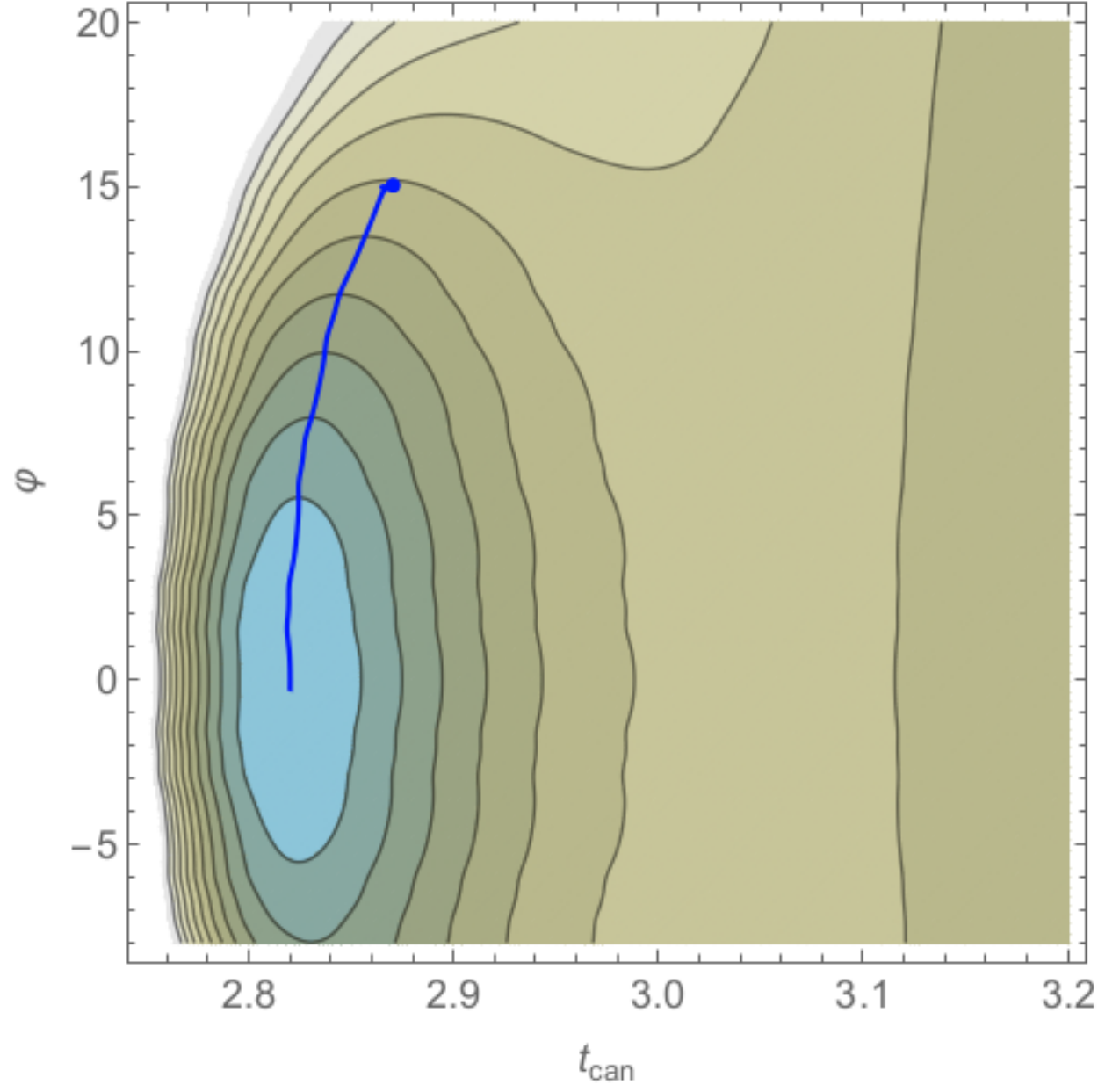}
    \label{fig:Per1cont}
 }\qquad\quad
 \subfigure[][~Effective scalar potential on the inflaton trajectory.]{
     \raisebox{7mm}{\includegraphics[width=.91\columnwidth]{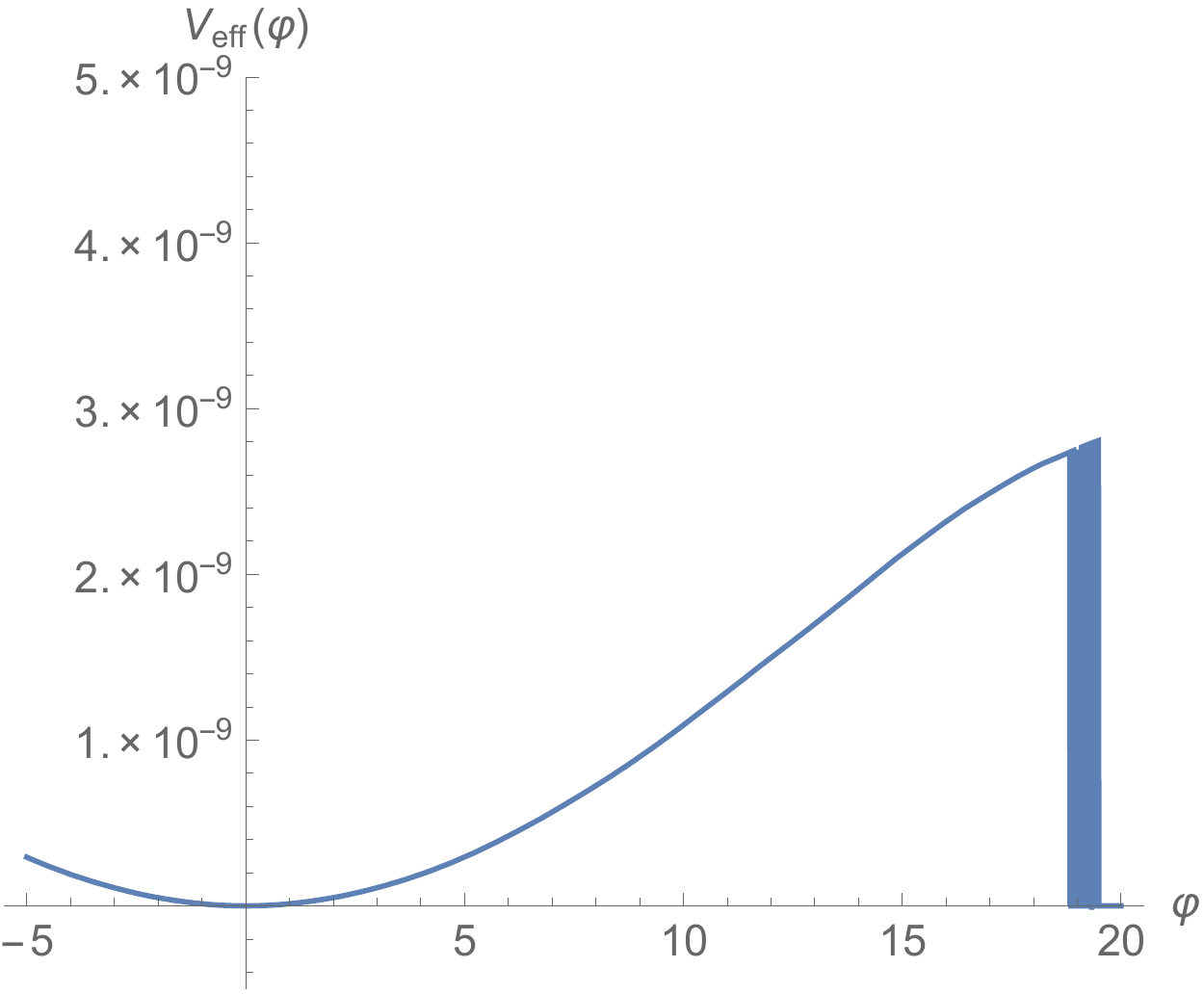}}
     \label{fig:Per1}
 } 
 \caption{\subref{fig:Per1cont} Full scalar potential for an example with $\vartheta_3$ and $\delta = 1/4 \pi^2$, with the axion of $T$ minimized dynamically. The blue line illustrates the flattest possible trajectory which evolves to the metastable Minkowski vacuum. Small modulations are clearly visible, leading to distinct features in the CMB. \subref{fig:Per1} Effective potential on the blue trajectory in Figure~\ref{fig:Per1cont}, with $t$ and its axion minimized via their field-dependent equation of motion.}
 \label{fig:ox3}
\end{figure*}
\begin{figure*}[t]
 \centering
\subfigure[][~Scalar potential in the $t$-$\varphi$-plane.]{
    \includegraphics[width=.88\columnwidth]{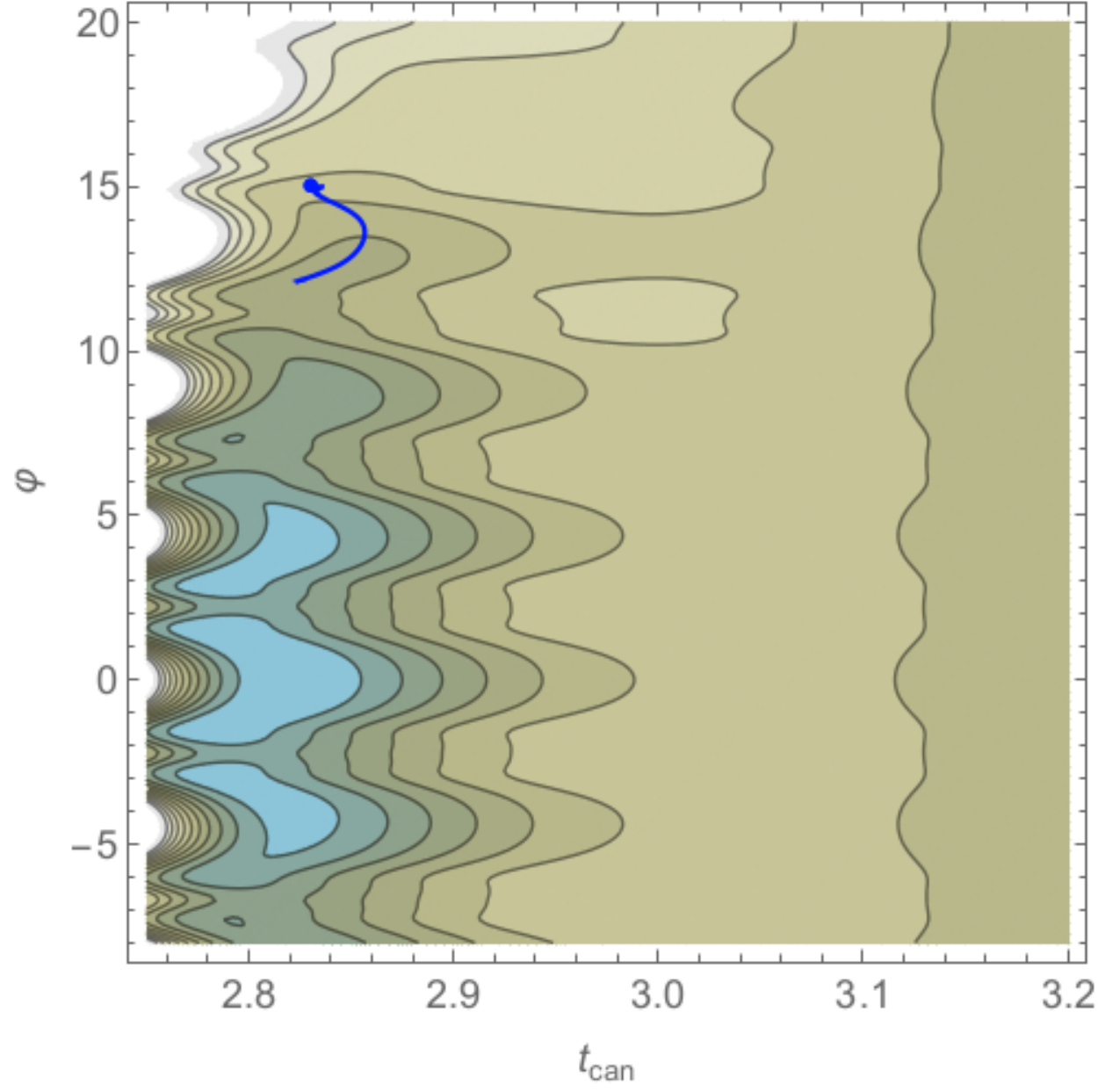}
    \label{fig:Per2cont}
 }\qquad\quad
 \subfigure[][~Effective scalar potential on the inflaton trajectory.]{
     \raisebox{7mm}{\includegraphics[width=.91\columnwidth]{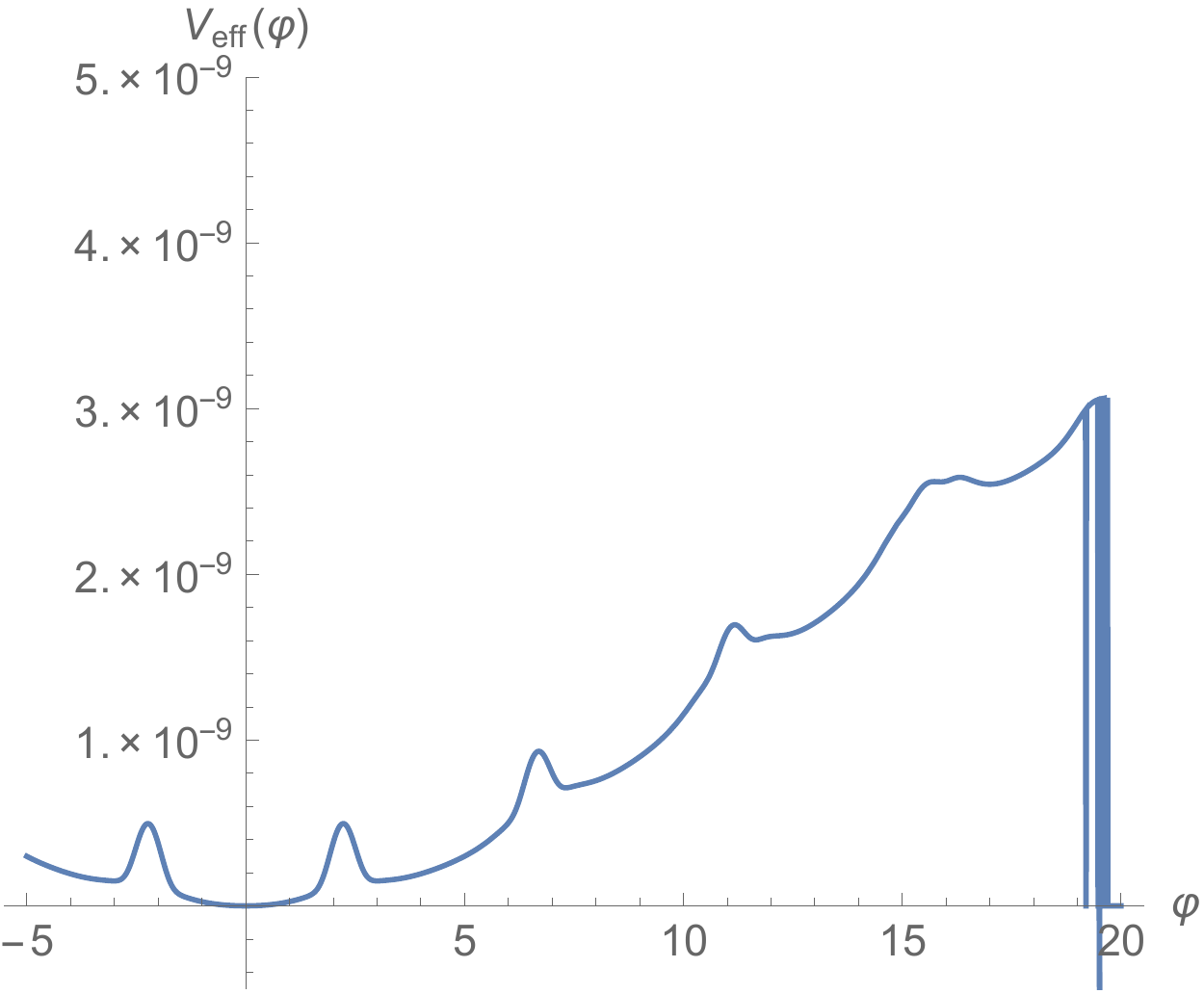}}
     \label{fig:Per2}
 } 
 \caption{\subref{fig:Per2cont} Full scalar potential for an example with $\vartheta_3$ and $\delta = 1/2$. The blue line illustrates that no possible trajectories exist in this case: the modulations are so large that the universe is trapped in false vacuum state, without inflating enough. \subref{fig:Per2} Effective potential on the blue trajectory in Figure~\ref{fig:Per2cont}, with $t$ and its axion minimized via their field-dependent equation of motion. The local minimum at $\varphi \approx 12$ in which the inflaton becomes trapped, when starting at $\varphi \approx 15$, is clearly visible.}
 \label{fig:ox4}
\end{figure*}

\subsubsection{Motivation}

The case of exponential one-loop corrections is in a sense a worst-case scenario: the D3-brane charge back-reacting on the geometry and the four-cycle to be stabilized are very close to each other. Since the change of the warped volume by induced D3-brane charge is a local effect, different situations may arise in other microscopic setups. For example, there may be situations where the one-loop Pfaffian on a four-cycle which is far away from the source of the backreaction only depends periodically on the brane-position $\Phi$. In the D7-brane inflation models considered here this can arise because the total D3-brane charge of the configuration must, of course, vanish. In the example of \cite{Ibanez:2014swa} this is realized by the simultaneous winding of a $\overline{\text{D7}}$-brane carrying units of induced $\overline{\text{D3}}$-brane charge. The geometric backreaction on a four-cycle far away from this brane system would not be sensitive to the full amount of D3-brane charge on the D7, but to a dipole system in which the effect of the opposite-charged branes partly cancels. In \cite{Flauger:2009ab,Flauger:2014ana} this idea was applied to one of the original setups of axion monodromy inflation.

\subsubsection{Theta functions}
\label{sec:Theta}

Also in this case, the explicit form of the one-loop Pfaffian strongly depends on the compactification one chooses. In \cite{Berg:2004ek,Berg:2004sj} it was shown that, in case of a toroidal manifold, when the four-cycle is insensitive to the pile-up of charge, the coefficient $A$ is proportional to $\vartheta$-functions. Thus, let us consider the following ansatz
\begin{align}
K &= -3 \log{(T + \bar T)} + \frac12 (\Phi + \bar \Phi)^2\,,\\
W &=  W_0 + \mu \Phi^2 + A_0 \vartheta_j(i\Phi,q)^\delta e^{-\alpha T}\,, \label{eq:PerW1}
\end{align}
where $\delta$ is again a positive constant proportional to $1/4 \pi^2$ that depends on the separation of the mobile D7/$\overline{\text{D7}}$-branes and the affected four-cycle. $q$ is, in our case, a complex number with $|q| < 1$ determined by vacuum expectation values of complex structure moduli, $q = e^{ i \pi \tau}$. The interesting cases are $j = 2$ and $j = 3$, since the two remaining theta functions are related by phase shifts. We use the following sum representations for the $\vartheta$-functions,
\begin{align}
\vartheta_2 (u,q) &= 2 q^{1/4} \sum_{n=0}^{\infty} q^{n(n+1)} \cos((2n+1)u)\,, \\
\vartheta_3 (u,q) &= 1 + 2 \sum_{n=1}^{\infty} q^{n^2} \cos(2 n u)\,.
\end{align}
Note that the frequency of the periodic functions is, throughout this section, given by the size of the transverse two-cycle that is wound by the D7-brane, which we take to be of $\mathcal O(1)$ in string units.

As before, we can test this setup for the possibility of large-field inflation, and constrain the parameter $\delta$. As a first example, let us choose $j = 3$, $q = 0.3 - 0.4 i$, $\delta = 1/4 \pi^2$, and the remaining parameters as in Section \ref{sec:Exp}. The corresponding contour plot can be found in Figure~\ref{fig:Per1cont}.

Evidently, the periodic one-loop Pfaffian introduces modulations to the potential, and subsequently to all possible slow-roll trajectories. In this case, the modulations are so small that they are hardly visible in a plot of the effective inflaton on the blue trajectory, cf.~Figure~\ref{fig:Per1}. Nevertheless, they leave an imprint in the CMB. For the case of sinusoidal modulations this has been studied many times in the literature, cf.~\cite{Kappl:2015esy} for an enlightening discussion and a list of references. In our case the shape of the modulations is governed by the shape of $\vartheta_3$, which in turn depends on the parameter $q$. In the example above, the predictions for $n_\text s$ and $r$ are well within the allowed regime, while the running of the spectral index is substantial and highly sensitive to the value of $\delta$. Moreover, due to the turns in the trajectory we cannot exclude that multi-field effect play a role in this setup. While this certainly opens a window to interesting CMB phenomenology and very distinctive features, a detailed study of the inflationary dynamics is beyond the scope of the present work.

However, we can still find an important constraint on the value of $\delta$. If the modulations are too pronounced, slow-roll inflation may be impaired or, even worse, the universe may become trapped in the wrong vacuum. We can illustrate an extreme example like that by choosing $\delta = 1/2$ and leaving the remaining parameters unchanged. The corresponding potential is illustrated in Figures~\ref{fig:Per2cont} and \ref{fig:Per2}.

\begin{figure*}[t]
 \centering
\subfigure[][~Scalar potential in the $t$-$\varphi$-plane.]{
    \includegraphics[width=.88\columnwidth]{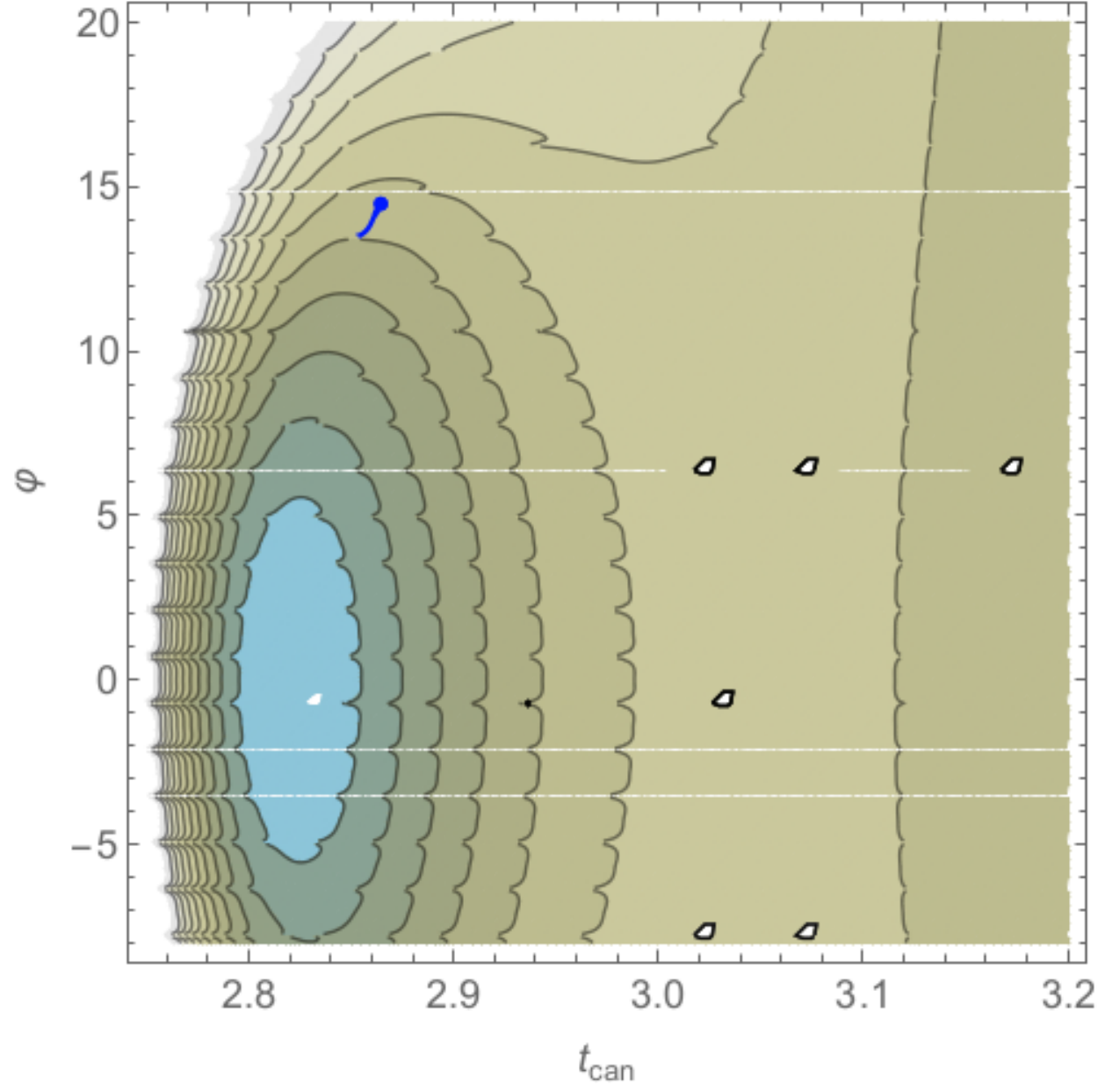}
    \label{fig:Per3cont}
 }\qquad\quad
 \subfigure[][~Effective scalar potential on the inflaton trajectory.]{
     \raisebox{7mm}{\includegraphics[width=.91\columnwidth]{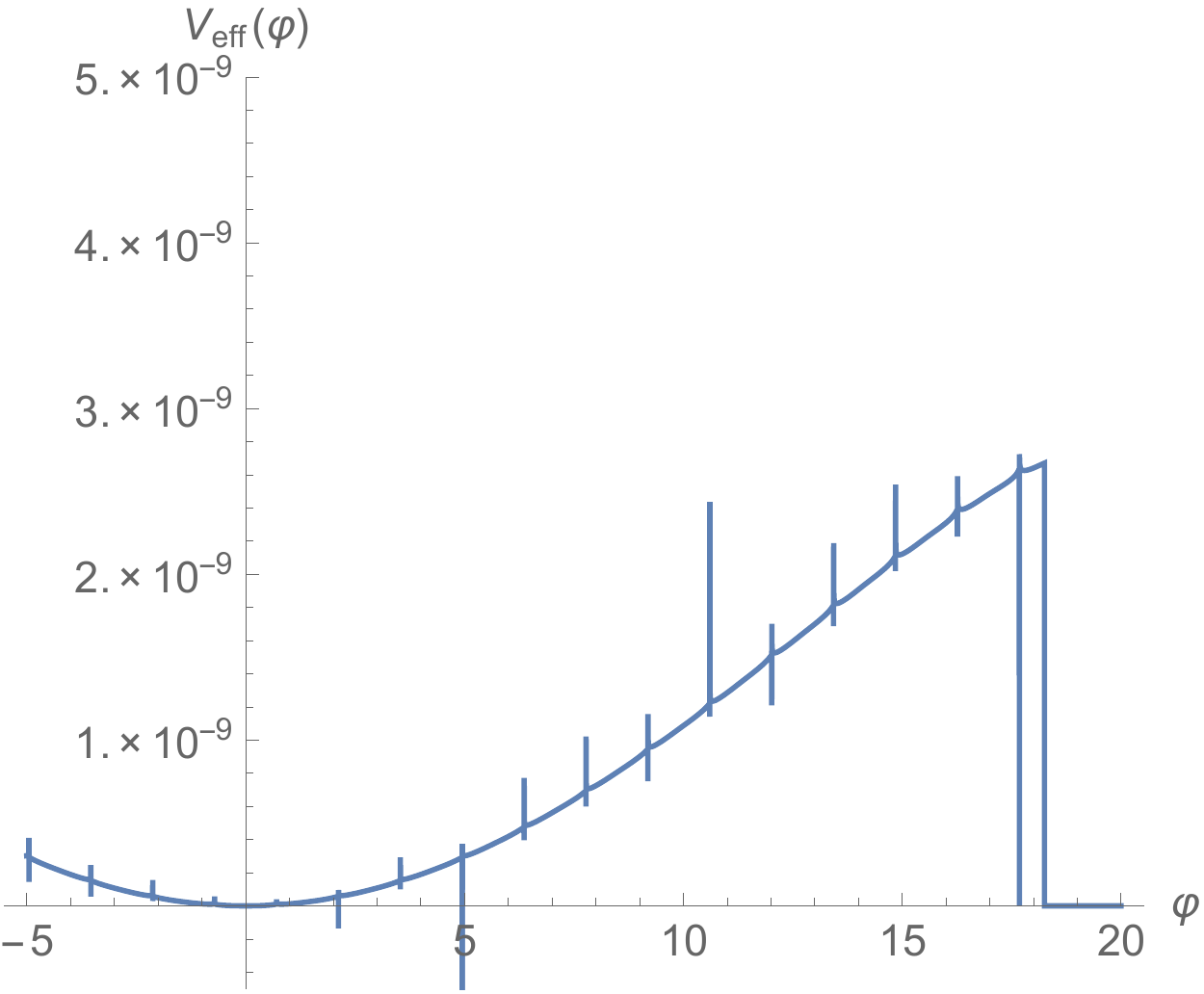}}
     \label{fig:Per3}
 } 
 \caption{\subref{fig:Per3cont} Scalar potential for an example with $\vartheta_2$ and $\delta = 1/4 \pi^2$. The small spikes in the contour plot indicate branch cuts of the theory. \subref{fig:Per3} Effective potential on a would-be trajectory in Figure~\ref{fig:Per3cont}, with $t$ and its axion minimized via their field-dependent equation of motion. The poles signal unphysical branch cuts. The difference in height of the visible spikes is due to numerical inaccuracies.}
 \label{fig:ox5}
\end{figure*}

A few comments on the case $j = 2$ are in order. The shape of $\vartheta_2$ is widely different from that of $\vartheta_3$ for the same values of $q$, so that modulations proportional to $\vartheta_2$ inevitably lead to branch cuts in the scalar potential. This is easily visualized by plotting the corresponding potentials in a parameter example with $j = 2$ and $\delta = 1/4 \pi^2$, a comparably small value. The results are shown in Figures~\ref{fig:Per3cont} and \ref{fig:Per3}.
As the blue trajectory indicates, the inflaton is immediately trapped in a false vacuum in a wrong branch of the theory. This is also clearly visible in the effective potential in Figure~\ref{fig:Per3}, which features multiple branch cuts. 

The form of the scalar potential in this case signals that the ansatz \eqref{eq:PerW1} with $j=2$ is not a low-energy effective theory of the microscopic setup we have in mind. For a large separation of the backreaction source and the perturbed four-cycle, the one-loop correction should be a small effect compared to the leading-order potential. Instead, the ansatz \eqref{eq:PerW1} with $j=2$ only describes the correct theory for very small values of $\varphi$ around the vacuum, until the first branch cut appears. While we cannot exclude that string theory constructions exist that produce this form of one-loop Pfaffian, they are at most compatible with small-field inflation models, like the one originally considered in \cite{Berg:2004ek,Berg:2004sj}.

\subsubsection{Trigonometric functions}
\label{sec:Trig}

Before closing this Section, we would like to remark that the qualitative features of the above models are captured very well by the leading-order expansions of the respective theta functions. Already the first non-trivial term in each expansion is sufficient to describe the physics qualitatively. Hence for $j = 3$ we may use the simplified ansatz
\begin{align}
W = W_0+\mu \Phi^2 + A_0 \left[3 + 2 q \cos(2 i \pi \Phi) \right]^\delta e^{-\alpha T}\,. \label{eq:PerW2}
\end{align}
In this case the shape of the modulations in the potential is sinusoidal. For $j = 2$, on the other hand, one may use the ansatz 
\begin{align}
W = W_0+\mu \Phi^2 +  A_0 \left[ 2 q^{1/4} \cos(i \pi \Phi) \right]^\delta e^{-\alpha T} \,. \label{eq:PerW3}
\end{align}
In this form the important difference between the two cases becomes very clear. While the argument of the square bracket in \eqref{eq:PerW2} is always positive, it fluctuates around zero in \eqref{eq:PerW3}. For $\delta \notin \mathbb N^+$ this leads to the observed branch cuts of the root function whenever the cosine is negative.

\section{Summary and Outlook}
\label{sec:Conclusion}

We have studied the viability of large-field inflation in string-effective supergravity models where the inflaton appears in the Pfaffian coefficient of the non-perturbative superpotential that stabilizes K\"ahler moduli. This leads to a potentially strong backreaction of the K\"ahler sector on the inflaton potential, in addition to the well-known backreaction from gravitational-strength interactions.

In cases where the Pfaffian arises from Euclidean branes or world-sheet instantons, the coefficient is a homogeneous polynomial in the inflaton field. Its degree influences the severity of the backreaction. We distinguish three cases and derive bounds on the EFT parameter $\delta$ which can, in principle, be computed in a given microscopic setup. We obtain our results numerically by tracking the minima of all fields dynamically along the inflaton trajectory. Remarkably, when the degree of the polynomial is odd or a multiple of four, the one-loop correction leads to a flattening of the potential, and inflation only fails for quite large values of $\delta$.

In cases where the Pfaffian arises from target-space instantons or corrections to the gauge-kinetic function, the coefficient is in many well-studied cases either an exponential or a periodic function. In the exponential case the backreaction of $t$ is such that the inflaton valley is tilted towards the exponentially steep small-volume region. Since the dependence on the inflaton is exponential, much smaller values of the parameter $\delta$ are needed for successful inflation. Controlling this backreaction is challenging in relaxion models but seems possible in chaotic inflation, which can do with smaller field excursions. If $\delta\sim 1/4\pi^2$ the backreaction is small enough to yield successful chaotic inflation, while for $\delta \gtrsim 0.1$ it becomes very challenging with reasonable parameter values. 

Finally, if the one-loop Pfaffian is a periodic function, for example a generalized theta-function, we find that the periodicity is imprinted on the inflaton trajectory in the form of modulations of the potential. This leads to a curved inflaton trajectory which might jeopardize single-field inflation and induce a running of the spectral index. In case of sinusoidal modulations such effects have been studied in detail in the literature, and they can lead to interesting signatures in the CMB. Since these modulations, or even more complicated ones, seem quite generic in string compactifications, we believe that the inflationary dynamics in setups with periodic one-loop corrections deserve a detailed study in the future. Furthermore, given the important role of the Pfaffian in moduli stabilization, it would be very interesting to study in which cases the Pfaffian vanishes at the end of inflation. In the models under consideration, this happens in the polynomial case if $\delta_0=0$ and in the periodic case if $\vartheta_1$ occurs. Interestingly, these apparently different cases seem to correspond to different parameterizations of the instanton moduli space \cite{Curio:2008cm,Curio:2009wn}. Their vanishing is linked to the occurrence of very symmetric points in the instanton moduli space \cite{Cvetic:2012ts} or to special constructions of the Calabi-Yau \cite{Berg:2004ek,Berg:2004sj,Beasley:2003fx}.

\section*{Acknowledgments}
We would like to thank Luis Ib\'a\~nez, Fernando Marchesano, Liam McAllister, John Stout, Irene Valenzuela, and Alexander Westphal for useful and interesting discussions. The work of FR is supported by the EPSRC grant EP/N007158/1 ``Geometry for String Model Building''. The work of CW is partially supported by the grants FPA2012-32828 from the MINECO, the ERC Advanced Grant SPLE under contract ERC-2012-ADG-20120216-320421, and the grant SEV-2012-0249 of the ``Centro de Excelencia Severo Ochoa" Programme.


\end{document}